\documentclass [11pt,letterpaper]{article}
\pdfoutput=1

\usepackage{jheppub}
\usepackage[mathscr]{eucal}
\usepackage{bm}
\usepackage{epsfig}
\usepackage[latin1]{inputenc}
\usepackage{float}
\usepackage{graphicx}
\usepackage{cancel}
\usepackage{mathrsfs}
\usepackage{amssymb}
\usepackage{amsfonts}
\usepackage{amsmath}
\usepackage{slashed}
\usepackage{hyperref}

\usepackage{caption}

\newcommand{\beq}{\begin{equation}}
\newcommand{\eeq}{\end{equation}}
\renewcommand\r[1]{(\ref{#1})}
\def\({\left(}  \def\){\right)}  \def\[{\left[}  \def\]{\right]}  \def\<{\langle}  \def\>{\rangle} 
\newcommand{\bew}{\begin{widetext}}
\newcommand{\ew}{\end{widetext}}

\title{Effects of Fluid Velocity Gradients on Heavy Quark Energy Loss}

\author{Mindaugas~Lekaveckas,}
\author{Krishna~Rajagopal}
\affiliation{Center for Theoretical Physics, Massachusetts Institute of Technology, Cambridge, MA 02139}
\emailAdd{lekaveck@mit.edu}
\emailAdd{krishna@mit.edu}

\preprint{MIT-CTP 4516}

\abstract{

We use holographic duality to analyze the drag force on, and consequent energy loss of, a heavy quark moving through a strongly coupled conformal fluid with non-vanishing gradients in its velocity and temperature. 
We derive the general expression for the drag force to first order in the fluid gradients. 
Using this general expression, we show that a quark that 
is instantaneously at rest, relative to the fluid, 
in a fluid whose velocity is changing with time
feels a nonzero force.
And, we show that for
a quark that is moving ultra-relativistically, the first order gradient ``corrections" become larger than the zeroth order drag force, suggesting that the gradient expansion may be unreliable in this regime. We illustrate the importance of the fluid gradients for heavy quark energy loss by considering a fluid with one-dimensional 
boost invariant Bjorken expansion as well as the
strongly coupled plasma created by colliding sheets of energy.

}


\date{\today}

\begin{document}

\maketitle

\section{Introduction and Summary}
\label{sec:intr}

The analysis of how a heavy quark moving through the strongly coupled liquid
quark-gluon plasma
produced in ultrarelativistic heavy ion collisions
loses energy and, subsequently, diffuses in the flowing plasma is of considerable
theoretical interest because experimentalists are developing the detectors
and techniques needed to
use heavy quarks as `tracers' or `probes' of the strongly coupled
liquid.
If one assumes that the interactions between the heavy quark
and the quark-gluon plasma are weak then perturbative
techniques originally formulated for energetic light 
quarks~\cite{Gyulassy:1993hr, Baier:1996kr, Baier:1996sk, Zakharov:1997uu}
can be employed to analyze heavy quark energy loss~\cite{Dokshitzer:2001zm}.

The discovery that the plasma produced in heavy
ion collisions is a strongly coupled liquid has prompted
much interest in the real-time dynamics of strongly coupled non-Abelian plasmas
and in the dynamics of heavy quarks therein.
Although it remains to be seen to what degree treating all aspects of the
dynamics of heavy quarks as strongly coupled is a good approximation,
this approach is certainly of value as a benchmark: thorough understanding of the physics
in this tractable setting can provide valuable qualitative insights.   
What makes these calculations tractable is holographic duality, which maps
questions of interest onto calculations done via a dual gravitational description
of the strongly coupled plasma and the heavy quark probe.  The simplest
theory in which these holographic calculations can be done is strongly
coupled ${\cal N}=4$ supersymmetric
Yang-Mills (SYM) theory in the large number of colors (large $N_c$) limit, whose
plasma with temperature $T$
is dual to classical gravity in a 4+1-dimensional spacetime
that contains a $3+1$-dimensional horizon with Hawking temperature $T$ and
that is asymptotically anti--deSitter (AdS) spacetime, with the heavy quark represented by a string
moving through this spacetime~\cite{Maldacena:1997re,Witten:1998qj,Karch:2002sh,Herzog:2006gh,Gubser:2006bz,CasalderreySolana:2006rq}.
The earliest work on heavy quark dynamics in the equilibrium plasma of
strongly coupled ${\cal N}=4$ SYM
theory~\cite{Herzog:2006gh,Gubser:2006bz,CasalderreySolana:2006rq} yielded
determinations of the drag force felt by a heavy quark moving through the static plasma
and the diffusion constant that governs the subsequent diffusion of the heavy quark 
once its initial motion relative to the static fluid has been lost due to drag.  The basic picture
of heavy quark dynamics that emerges, with all but the initially most energetic heavy quarks
being rapidly slowed by drag and then becoming tracers diffusing within the (moving) fluid, is qualitatively
consistent with early experimental investigations~\cite{Adare:2006nq}.  For a 
review, see Ref.~\cite{CasalderreySolana:2011us}.
Subsequently, the holographic calculational techniques were generalized to any static plasmas
whose gravitational dual has a 4+1-dimensional metric that depends only on the
holographic (i.e.~`radial') coordinate in Ref.~\cite{Herzog:2006se} and heavy quark energy loss
and diffusion has by now been investigated in the equilibrium plasmas of many gauge theories
with gravitational duals~\cite{Caceres:2006dj, Caceres:2006as, Matsuo:2006ws, Nakano:2006js, Talavera:2006tj, Gubser:2006qh, Bertoldi:2007sf, Liu:2008tz,Gursoy:2009kk, HoyosBadajoz:2009pv, Bigazzi:2009bk, NataAtmaja:2010hd, Chernicoff:2012iq, Fadafan:2012qu}. 

More recently, in Ref.~\cite{Chesler:2013cqa} we and a coauthor have calculated how the drag force
and energy loss rate of a heavy quark moving through the far-from-equilibrium
matter present just after a collision compares to that in strongly coupled plasma close
to equilibrium.  We studied the energy loss of a heavy quark moving through the debris
produced by the collision of planar sheets of energy in strongly coupled SYM
theory introduced in Ref.~\cite{Chesler:2010bi} and analyzed there 
and in Refs.~\cite{Casalderrey-Solana:2013aba,Chesler:2013lia}.
The matter produced in these collisions is initially far from equilibrium
but then rapidly hydrodynamizes: its expansion and
cooling is described well by viscous hydrodynamics after a time $t_{\rm hydro}$
that is at most around $(0.7-1)/T_{\rm hydro}$, where $T_{\rm hydro}$
is the effective temperature defined from the fourth root of the energy
density at the hydrodynamization time $t_{\rm hydro}$.  
In Ref.~\cite{Chesler:2013cqa} we computed the drag force on a heavy
quark moving through the initially far-from-equilibrium matter and the 
subsequent hydrodynamic fluid.  We compared our results to 
what the drag force would have been in an equilibrium fluid
with the same instantaneous energy density, and found that
there is no dramatic ``extra'' energy loss in 
the far-from-equilibrium matter.  However, even at late times when
the expansion of the fluid is well-described by viscous hydrodynamics
we found deviations between the actual drag force and what the drag
force would have been in a spatially homogeneous equilibrium fluid 
with the same energy density. That is, we found that the gradients
in the actual fluid {\it do} affect the drag force felt by the heavy quark
moving through the fluid.  Our goal in the present paper is a thorough
investigation of the effects of gradients in the temperature and velocity
of the fluid, up to first order, on the drag force.

We begin by computing the drag force on a heavy quark moving through 
a fluid whose own motion is only in one direction, which we shall take
to be the $z$-direction.  If we denote the fluid $4-$velocity by $u^\mu$ 
then at this stage the only gradients that we are considering 
are $\partial_z u^z$ and $\partial_t u^z$ as well as $\partial_z T$ and $\partial_t T$.
Throughout this paper, we shall only work to first order in spatial gradients
and time derivatives of the fluid temperature and velocity.  The gravitational dual for
a slowly changing fluid, including the effects of first order derivatives but neglecting
higher derivatives, was first obtained in Ref.~\cite{Bhattacharyya:2008jc}, where
Einstein's equations in the 4+1-dimensional gravitational theory were solved to
first order in gradients in the boundary coordinates and exactly in the radial
direction.  In Section \ref{sec:string_grav} we describe this metric, for the case where the fluid motion is
only in one direction, and then in Section \ref{sec:string_quark} we introduce a heavy quark, described in the
gravitational theory by a string.  The endpoint lives at the boundary of the
AdS space, where it follows the trajectory of the heavy quark of interest.
We shall assume that it is being dragged at some constant velocity $\vec \beta$, which may or may not be parallel to the direction of motion of the fluid.\footnote{  
We shall work throughout in the heavy quark mass, {\it i.e.}~$M\rightarrow\infty$, limit and we shall assume throughout that the quark is being pulled at constant velocity $\vec\beta$ by some external force. We leave to future work the consideration of the case where there is no external force meaning that 
a quark with finite $M$ would decelerate under the influence of the force exerted on it by the fluid. }
%\textcolor{red}{Since we consider quark to be heavy, the approximation that quark is moving with constant velocity is consistent --  if the quark is let to slow down, the velocity would change very slowly due to inertia.}
In Section \ref{sec:string_quark}, with the gravitational metric describing
the fluid in hand, we formulate and solve the string equations of motion,
which is to say that we calculate the shape of the string attached to the
heavy quark, including effects of fluid gradients up to first order.
In Section \ref{sec:forces_RF} we use the string profile to calculate the flux of momentum
down the string, which
determines the drag force
on the heavy quark.  We first do the calculation in the fluid rest frame and then
boost the result to any frame in Section \ref{sec:forces_gen}.
In Section \ref{sec:forcegen} we generalize to the case in which
the fluid has an arbitrary velocity and in which any of the gradients $\partial_\alpha u^\beta$
can be nonzero.

By analyzing the case in which the quark is moving with an ultrarelativistic
velocity relative to the fluid we find indications that our results
may not be valid in the limit in which the Lorentz factor $\gamma$
of the quark velocity is large, even if the quark mass $M\rightarrow\infty$
limit has been taken first and even if fluid gradients are small.
We find that in the $\gamma\rightarrow\infty$ limit
the ``correction'' to the drag force that is first order in
fluid gradients is larger than the leading (i.e. zeroth order) 
term by a factor that is ${\cal O}(\gamma^{1/2})$.
This suggests that the gradient expansion may not be valid in this regime. That is,
 even if higher order gradients are small enough that they are not important
in describing the fluid motion itself their effects on the drag force may become important 
at large enough $\gamma$.

In Section~\ref{sec:applications},
we consider three consequences of our general result for the first order effects of fluid
gradients on the drag force exerted by the fluid on the heavy quark.  First, we point
out that even if the quark has no velocity relative to the fluid the drag force on it
is nonzero as long as the time derivative of the fluid velocity is nonvanishing.
Next, we consider two explicit examples of a fluid
whose motion is only in one direction.
First, we analyze the drag force on a heavy quark moving through a fluid that is
undergoing boost-invariant expansion in the $z$-direction, \`a la Bjorken.
We show that even though there are gradients in this fluid, 
a quark that is moving along with the fluid feels no drag force.
A quark whose velocity includes a component perpendicular
to the direction of motion of the fluid feels a drag force
that is affected by the fluid gradients.
And, last of all, we return to the colliding sheets of energy density
that motivated our investigation, showing that our general expression
for the effects of fluid gradients on the drag force to first order does a good job of explaining the explicit results obtained in
Ref.~\cite{Chesler:2013cqa}, in many cases quantitatively and qualitatively in
all cases, even those where the results of Ref.~\cite{Chesler:2013cqa} appear counter-intuitive. In Section~\ref{sec:dis} we look ahead to new directions whose investigation is motivated by our results.

\section{Hydrodynamic fluid and a heavy quark moving through it}
\label{sec:string}

\subsection{Gravitational description of a moving fluid}
\label{sec:string_grav}

We begin with a brief description of the dual gravitational description of the
stress-energy tensor for the conformal fluid
of strongly coupled ${\cal N}=4$ SYM theory at nonzero temperature, undulating
in some generic way according to the laws of hydrodynamics.  We shall work only to first order
in fluid gradients.    In order to keep our expressions
tractable on a first pass through the calculation, we shall then specialize to the case of a  fluid that
fills 3-dimensional space but that moves only along a single
axis, flowing in some generic way along the $z$-direction.  (Toward the end of Section \ref{sec:forces} we shall lift this
restriction, returning there to the case of generic hydrodynamic motion in 3+1 dimensions, still working
only to first order in gradients which is to say still assuming that the spatial and temporal variation of the 
thermodynamic variables and the fluid velocity occur only on length and time scales that are much longer
than $1/T$, with $T$ the fluid temperature.)

The stress-energy tensor for 
the conformal fluid of ${\cal N}=4$ SYM theory flowing
hydrodynamically in $3+1$-dimensions with a temperature $T$
and $4$-velocity $u^\mu$ that vary as functions of
space and time
is given
to first order in gradients by
\beq
\frac{8\pi^2}{N_c^2} T^{\mu\nu} = \frac{1}{b^4} \(\eta^{\mu\nu} + 4 u^\mu u^\nu\) - \frac{2}{b^3} \sigma^{\mu\nu}, 
\label{Tmunu}
\eeq
where $b \equiv 1/(\pi T)$ is the inverse temperature, $\eta^{\mu\nu} = \text{diag} (-1,1,1,1)$
is the Minkowski metric, and 
\beq
\sigma^{\mu\nu} = \frac{1}{2}P^{\mu\alpha} P^{\nu\beta} (\partial_\alpha u_\beta + \partial_\beta u_\alpha) - \frac{1}{3} P^{\mu \nu} \partial_\alpha u^\alpha
\label{sigmamunu}
\eeq
is
the symmetric tensor encoding the first order contributions of fluid gradients, 
with the projectors transverse to $u^\mu$ defined by $P^{\mu\nu} \equiv \eta^{\mu\nu} + u^\mu u^\nu$.   
With our metric conventions, $u^\mu$ is normalized such that $u^\mu u_\mu=-1$.
The stress-energy tensor (\ref{Tmunu}) describes a fluid whose equation of
state is $P=\varepsilon/3$, where $P$ and $\varepsilon$ are its pressure
and energy density respectively, and whose shear and bulk viscosities
are given by $\eta=s/(4\pi)$ and $\zeta=0$, where $s$ is the entropy density
of the fluid.  $P=\varepsilon/3$ and $\zeta=0$ follow just from conformal invariance.
$4\pi\eta/s = 1$ for the fluid in any non-Abelian gauge theory with a 
dual gravitational description, in the strong coupling 
and large-$N_c$ limit~\cite{Policastro:2001yc,Kovtun:2003wp,Buchel:2003tz,Kovtun:2004de}.
Note that the
stress-energy tensor depends on symmetric combinations of the fluid gradients $\partial_\alpha u_\beta$ and is independent of the fluid vorticity 
\beq
\tilde\omega^\mu \equiv  \frac{1}{2} \epsilon^{\mu \nu \alpha \beta} u_\nu \partial_\alpha u_\beta\,,
\label{vorticity_definition}
\eeq
because
the underlying microscopic theory does not violate time-reversal or parity symmetry.
The vorticity will nevertheless play a role in our considerations later.
Hydrodynamics is the statement that 
the fluid variables satisfy energy and momentum conservation, 
\beq
\partial_\mu T^{\mu\nu} = 0\ .
\label{HydrodynamicEquations}
\eeq
It is easy to see that, to first order in gradients, the hydrodynamic equations (\ref{HydrodynamicEquations})
determine the spatial and temporal variation of the temperature, or the inverse
temperature $b$, uniquely in terms of the spatially and temporally varying fluid velocities:
\beq
\partial_\mu b = b \( u^\nu \partial_\nu u_\mu - \frac{1}{3}  u_\mu \partial_\alpha u^\alpha \).
\label{partialb}
\eeq
We shall use this relation below.

The dual gravitational description of the fluid with stress-energy tensor (\ref{Tmunu}) was
obtained in Ref.~\cite{Bhattacharyya:2008jc}.  (These authors worked to second order
in gradients. We shall quote their results only to first order.)
Upon introducing a bookkeeping parameter $\epsilon$ that we shall use to count
powers of gradients and that we shall in the end set to $\epsilon=1$, 
the $4+1$-dimensional metric in the dual gravitational
description of the fluid takes the form
\beq
ds^2 = \(G_{MN}^{(0)} + \epsilon\, G_{MN}^{(1)}\)dX^M dX^N
\label{GMN}
\eeq
where  $X^M\equiv (x^\mu, r)$. 
The first term in \r{GMN} is given by
\beq
\begin{split}
&G_{MN}^{(0)}dX^M dX^N = -2u_\mu dx^\mu dr - r^2 f(br) u_\mu u_\nu dx^\mu dx^\nu + r^2 P_{\mu \nu} dx^\mu dx^\nu
\end{split}
\label{GMN0gen}
\eeq
where $f(x) \equiv 1 - 1/x^4$.   If we set $u^\mu=(1,0,0,0)$ everywhere, the metric (\ref{GMN0gen}) describes
a static AdS black brane with an event horizon at $r=1/b$ 
and with the AdS boundary 
located at $r=\infty$.  This is the gravitational dual of the static ${\cal N}=4$ SYM plasma
in equilibrium with a uniform and constant temperature $T=1/(\pi b)$.  The coordinates
$X^M$ that we are using to describe the spacetime,
chosen in such a way that the metric has no $dr^2$ term and
has no singularities at $r=1/b$, are referred to 
as in-falling Eddington-Finkelstein coordinates.
With a generic choice of $u^\mu$, varying as a function of $x^\mu$,
at any given $x^\mu$ the metric (\ref{GMN0gen}) is obtained by boosting
the AdS black brane metric by the boost that takes you from the instantaneous fluid
rest frame, where $u^\mu=(1,0,0,0)$,  to the frame in which $u^\mu$ takes on the value of interest.
The metric (\ref{GMN0gen}) is therefore often said to describe a boosted black brane, but
it is important to remember that $b$ and $u^\mu$ are in fact varying.  
It describes a black brane whose horizon is undulating, as is its entire metric.
Note that although $r=1/b$ is the horizon of the static black brane, once $1/b$ is undulating the global event horizon of the metric \r{GMN} need no longer be located at $r=1/b$.  Gradient  corrections to the position of the event horizon have been computed in Ref.~\cite{Bhattacharyya:2008xc}.
The metric (\ref{GMN0gen}) is
the zeroth approximation to the gravitational dual of the moving fluid; it would be 
a complete description if gradients made no contribution to the fluid stress-energy
tensor, which is to say if the fluid were an ideal fluid with zero shear viscosity.

The second term in the metric \r{GMN} is the dual gravitational description
of the contribution of first order gradients in $u^\mu$ and $b$ to the fluid
stress-energy tensor. It is given by~\cite{Bhattacharyya:2008jc}
\beq
\begin{split}
&G_{MN}^{(1)}dX^M dX^N = 2 r^2 b\, F(br) \sigma_{\mu\nu} dx^\mu dx^\nu + \frac 2 3 r \,u_\mu u_\nu \partial_\lambda u^\lambda dx^\mu dx^\nu - r \,u^\lambda \partial_\lambda (u_\mu u_\nu) dx^\mu dx^\nu
\end{split}
\label{GMN1gen}
\eeq
where
\beq
F(x) \equiv \frac{1}{4} \[ \log \( \frac{(1+x)^2(1+x^2)}{x^4} \) - 2\arctan x + \pi \].
\eeq
%where because of the `background field' gauge choice, 
We are working in a gauge in which
$G^{(1)}_{M r} = 0$. The metric \r{GMN1gen} is a good approximation to the gravitational
dual of the hydrodynamics of the  flowing conformal fluid 
%gravity dual for the conformal fluid 
as long as the length scale $L$ over which $b$ and $u^\mu$ vary satisfies $L\gg b$. 

In the next subsection, we will compute the profile of the string 
that hangs ``down'' into the bulk metric $G_{MN}$ from the heavy
quark.  
To determine the profile of the string at the time $t$ at which the
heavy quark is located at a particular position $\vec x$,
it will prove convenient to do the calculation 
in the frame in which the fluid is at rest at $\vec x$ at
the instant of time $t$, which is to say the frame
in which $u^\mu(\vec x,t)=(1,0,0,0)$.
In making this choice we do not lose any generality 
since we can of course later boost the result of our
calculation to any frame that we like.
In order to do the calculation in the instantaneous 
fluid rest-frame it will be helpful to have the metric
$G_{MN}$ in this frame, which we obtain by setting $u^{\mu}=(1,0,0,0)$ 
in (\ref{GMN0gen}) and (\ref{GMN1gen}), keeping derivatives of $u^\mu$.
%Since we will only work to first order in fluid gradients, 
At the same time, 
since we will calculate 
the drag force on a heavy quark located at $x^\mu=0$ 
we expand $b(x^\nu)$ and $u^\mu(x^\nu)$
around $x^\nu=0$ in (\ref{GMN0gen}), 
keeping only terms that are  first order 
in their gradients.  Combining (\ref{GMN0gen}) and (\ref{GMN1gen}), the metric
then takes the form
\beq
\begin{split}
&G_{MN}dX^M dX^N = 2 dt dr - r^2 f(b r) dt^2  + r^2 dx_i dx_i \\
&+ \epsilon \(- 2 x^\mu \partial_\mu u_i dr dx^i - 2 x^\mu \partial_\mu u_i r^2 (1 - f(br))dt dx^i - 4 \frac{x^\mu \partial_\mu b}{b^5 r^2} dt^2 \right. \\
&\left. + 2 b r^2 F(b r) \sigma_{ij} dx^i dx^j  + \frac{2}{3} r \partial_i u_i dt^2 + 2r \partial_t u_i dt dx^i \)\ ,
\end{split}
\label{GMNgen}
\eeq
%where we used that $\partial_\mu u^0 = 0$ in the fluid rest frame.
which is the form that we shall need.

We shall begin by doing the calculation for a fluid that is only moving
in one direction, that we shall choose to be the $z$-direction.  In this case,
when we boost to the instantaneous rest-frame in which $u^\mu(\vec x,t)=(1,0,0,0)$
the only non-vanishing gradients are
%We will concentrate on the case where only one component of the fluid velocity $u^3$ has non-vanishing gradients along the time and $z-$direction
\beq
\partial_t b(t,z)\neq 0, \quad \partial_z b(t,z)\neq 0, \quad
\partial_{t} u^3(t,z) \neq 0, \quad \partial_{z} u^3(t,z) \neq 0 , 
\label{NonzeroGradients}
\eeq
with
\beq
\partial_\mu u^\perp = \partial_\perp u^\nu = \partial_\perp b = 0. % \partial_x u^\nu = \partial_y u^\nu = \partial_x b = \partial_y b = 0.
\label{VanishingGradients}
\eeq
This fluid configuration will not be sufficient for us to determine the drag force in
the most general configuration, in particular because in this configuration the
fluid has zero vorticity.  However, we shall see by the end of Section \ref{sec:forces}
that it suffices to get us most
of the way.
%With this choice, the gradients have zero vorticity, $\nabla \times \vec u = 0$ and provides a good example of a fluid. 
%As we will see in the next section, such choice of fluid gradients together with the choice of drag velocities suffices to determine drag force corrections in any frame.
Upon making this simplifying assumption,  conservation of the stress-energy tensor \r{partialb} takes on
the particularly simple form
\beq
%\left \{
\begin{split}
3\partial_t b(t,z) &= b(t,z) \partial_z u^3(t,z), \\
%\partial_x b &= 0\\
%\partial_y b &= 0 \\
\partial_z b(t,z) &= b(t,z) \partial_t u^3(t,z).
\end{split}
%\right.
\label{eos}
\eeq
%As we will see in section \ref{sec:forcegen}, perpendicular gradient has to be included to obtain drag force in a fluid with any velocity and drag velocity profiles, but the calculation is easy to generalize in such case.
We will return to the consideration of a general fluid configuration
only in Section~\ref{sec:forcegen}.

\subsection{Gravitational description of a moving heavy quark}
\label{sec:string_quark}

The dual gravitational string of a quark with mass $M$ 
at the spacetime point $x^\mu=0$ and moving with velocity $\vec \beta$
is, in the $M\rightarrow \infty$ limit, a string whose endpoint is at $x^\mu=0$ moving with velocity
$\vec\beta$ along the AdS boundary, namely at $r\rightarrow\infty$.
The dynamics of the string is described by the Nambu-Goto action
%To describe the dynamics of the string, we will use Nambu-Goto action, 
\beq
S_{\rm NG} = - \frac{\sqrt{\lambda}}{2\pi} \int d\tau d\sigma \sqrt{-g(\tau,\sigma)}
\label{SNG}
\eeq
where the string tension is $\frac{\sqrt{\lambda}}{2\pi}$, where $\lambda = g^2 N_c$ is the 't Hooft coupling, 
and where $g(\tau,\sigma) = \det g_{ab}(\tau,\sigma)$ with $g_{ab}(\tau,\sigma)$ 
the induced metric on the world-sheet, namely
\beq
\begin{split}
g_{ab}(\tau,\sigma) &= G_{MN} \partial_a X^M(\tau,\sigma) \partial_b X^N(\tau,\sigma).
\end{split}
\eeq
%To simplify the discussion as well as equation of motion we employ static gauge for the string world-sheet parametrization, so that
We shall parametrize the string world-sheet in such a way that
\beq
\begin{split}
t(\tau, \sigma) &= \tau, \\
r(\tau, \sigma) &= \sigma.
\end{split}
\label{static_gauge}
\eeq
We shall assume that the string is being dragged along with a constant
velocity $\vec\beta$. Because we are treating the case where the 
fluid motion is only in the $z$-direction, without loss of generality
we can choose $\vec\beta=(\beta_x,0,\beta_z)$.
We can think of the motion of the quark 
as being due to a force exerted
on it by some electric field, with respect to which
the quark is charged.  Our task is to determine the
force required to drag the quark, working to leading order in 
the fluid gradients.  The first step in the calculation, which we 
shall carry out in this section, is the determination of the string
profile, again to leading order in fluid gradients.

We denote the string profile to first order in gradients by
% at leading order can be generally written as
%\beq
%\begin{split}
%x(\tau, \sigma) &= x_0(\tau, \sigma) + \epsilon b \rho_x (\tau, \sigma) \\
%z(\tau, \sigma) &= z_0(\tau, \sigma) + \epsilon b \rho_z (\tau, \sigma)
%\end{split}
%\eeq
%and for simplicity we can choose $y(\tau, \sigma) = 0$. 
\beq
\vec x(\tau, \sigma) = \vec x_0(\tau, \sigma) + \epsilon\, \vec x_1(\tau,\sigma) % b \vec \rho (\tau, \sigma) 
\label{x_profile}
\eeq
where $\vec x_0(\tau,\sigma)$ is the string profile in the case of an equilibrium 
fluid with a constant temperature that is moving with some uniform velocity, which is to say in the absence of any gradients in the fluid velocity
or $b$.  In the instantaneous fluid rest-frame in which we are working this
means that $u^\mu=(1,0,0,0)$ and all gradients vanish.  This ``trailing string'' solution
was first obtained in Refs.~\cite{Herzog:2006gh,Gubser:2006bz} and is given by
%for constantly boosted fluid, for general fluid velocity given by 
\beq
\vec x_0(\tau, \sigma) = \vec \beta \left(  \tau - b    \(\tan^{-1}(b \sigma) - \frac{\pi}{2} \) \right)\ ,
\label{x0gen}
\eeq
%and he vector function $\vec \rho(\tau,\sigma)$ 
where we note that at $\sigma\rightarrow\infty$ the endpoint of the string follows the
trajectory of the heavy quark. We will need to differentiate $\vec x_0$, and to that
end we need to keep track of how it depends on $u^\mu$, namely
\beq
\vec x_0(\tau, \sigma) = \vec \beta \tau - b\(u^0 \vec \beta - \vec u\) \(\tan^{-1}(b \sigma) - \frac{\pi}{2} \)\ .
\eeq
The function $x_1(\tau,\sigma)$ in (\ref{x_profile})
encodes the corrections to the zeroth order profile $\vec x_0(\tau,\sigma)$ due to fluid gradients,
up to first order in those gradients.  It must vanish in the $\sigma\rightarrow\infty$ limit.
Our task in the remainder of this section is to calculate $x_1(\tau,\sigma)$.

The equations of motion for the string are obtained by extremizing 
the Nambu-Goto action with respect to the function $\vec x(\tau,\sigma)$.
To zeroth order we obtain $\vec x_0(\tau,\sigma)$.  The function $\vec x_1(\tau,\sigma)$ is determined from
%world-sheet coordinates $\tau$ and $\sigma$ or, equivalently, we can vary the action with respect to the unknown function $\vec \rho(\tau,\sigma)$, similarly as it was done in the original formulation of the drag force in the static fluid \cite{Gubser:2006bz},
\begin{align}
\partial_\tau \( \frac{\delta \mathcal{L}}{\delta (\partial_\tau \vec x_1) } \) + \partial_\sigma \( \frac{\delta \mathcal{L}}{\delta (\partial_\sigma \vec x_1) } \) = \frac{ \delta \mathcal L}{\delta \vec x_1}\ .
\label{eomsGen}
\end{align}
%By construction, at leading order the equations of motion are satisfied because we are perturbing the static solution $\vec x_0$ given in \r{x0gen}. 
Since the terms that are linear in gradients in (\ref{GMNgen}) arose either
directly from (\ref{GMN1gen}) or via expanding (\ref{GMN0gen}) to first order about
$x^\mu=0$, the terms in (\ref{eomsGen}) that are first order
in gradients can depend on time at most linearly, meaning
that $x_1(\tau,\sigma)$ takes the form
%so the function $\vec \rho(\tau,\sigma)$ can be decomposed as
\beq
\vec x_1 (\tau, \sigma) = b\, \tau\, \vec h(\sigma)  + b\, \vec g(\sigma)\ ,
\label{rho_dec}
\eeq
with $\vec h(\sigma)$ and $\vec g(\sigma)$ being dimensionless functions
that we must determine, that both have only $x$- and $z$-components,
and that both vanish at $\sigma\rightarrow\infty$.

The terms in the Euler-Lagrange equation (\ref{eomsGen}) that are
proportional to $\tau$ depend only on $\vec h(\sigma)$, not on $\vec g(\sigma)$.
It is in fact possible to guess the form of $\vec h(\tau)$.  However, determining it
by explicit solution is instructive, so we shall follow that route.
Integrating the Euler-Lagrange equations for $\vec h(\sigma)$ once yields
\beq
\begin{split}
h_x'(\sigma) &= \frac{c_{hx}\gamma^2 (b^2 \sigma^2 + 1)^2 - 2\beta_x D_t b (\gamma^2 + 2b^2\sigma^2 +1)}{(\gamma^2 - b^2 \sigma^2) (b^2 \sigma^2 + 1)^2}, \\
h_z'(\sigma) &=  \frac{1}{\left(b^4\sigma^4-\gamma ^2\right)} \times \\
&\( \frac{b  \left(b^4\sigma^4+2 b^2\sigma^2-\gamma ^2+2\right) D_t u^3}{\left(b^2\sigma^2+1\right)} + \frac{2 \beta_z\left(2 b^2\sigma^2+\gamma^2+1\right) D_t b}{\left(b^2\sigma^2+1\right)^2} - \gamma^2 c_{hz} \),
\label{hprime}
\end{split}
\eeq
%\beq
%h_z'(\sigma) = \frac{\frac{\left(\left(b^2 \sigma ^2+1\right) \left(b \beta^2 D_t u^3 + 4 \beta_z \left(\beta^2-1\right) D_t b \right)+b \left(b^2 \sigma ^2+1\right)^3 \left(\beta^2-1\right) D_t u^3-2 \beta_z \beta^2 D_t b \right)}{\left(b^2 \sigma ^2+1\right)^2}+c_{hz}}{b^4 \sigma ^4 \left(\beta^2-1\right)+1},
%%\end{split}
%\eeq
%
where by $'$ we mean $d/d\sigma$ and where 
\beq
D_t \equiv \partial_t + \beta^i \partial_i = \partial_t + \beta_z \partial_z
\label{convective_derivative}
\eeq
is the convective derivative along the path of the quark, with
the second equality valid here because the only nonzero gradients are in the $z$-direction, 
%$D_t u^3 \equiv \partial_t u^3 + \beta_z \partial_z u^3$, $D_t b \equiv \partial_t b + \beta_z \partial_z b$ denoting the total derivative of $u^3$ and $b$ along the path of the quark, 
where $\gamma = (1-\vec \beta^2)^{-1/2}$ is the Lorentz factor for the heavy quark,
and where
$c_{hx}$ and $c_{hz}$ are integration constants that we must now fix.
The expressions for $\vec h'(\sigma)$ have a first order pole at the radial position 
\beq
\sigma = \frac{\sqrt \gamma}{b}\ ,
\eeq
which in the case of the static fluid is identified as the location of the worldsheet horizon $\sigma_{\rm ws} \equiv  \sqrt{\gamma}/b$ that arises in the calculation of $\vec x_0(\tau,\sigma)$ in a static fluid~\cite{Herzog:2006gh,Gubser:2006bz}.

We have found that the position on the worldsheet where
the integration constants are fixed, $\sigma= \sqrt{\gamma}/b$, is the same as it would be in a static homogeneous fluid with the same instantaneous temperature.   
This means that our results disagree with those of 
Refs.~\cite{Abbasi:2012qz,Abbasi:2013mwa}: those authors
assumed that the influence of fluid gradients on the drag force
could be described via a dependence of this radial position
on the fluid gradients. We now see by explicit calculation that,
at least to first order, there is no such dependence.  And, indeed
our results for the drag force 
that we shall present in Section \ref{sec:forces} do differ from those in Refs.~\cite{Abbasi:2012qz,Abbasi:2013mwa}.

As in the static fluid  calculation of Refs.~\cite{Herzog:2006gh,Gubser:2006bz}, 
in order to obtain a regular string profile across the world-sheet horizon
we must choose the
integration constants in (\ref{hprime})  in such a way that the numerators on the right-hand sides
of (\ref{hprime}) vanish at the same $\sigma$ at which the denominators vanish, i.e. at the
world-sheet horizon.
%indicating that the string would not cross the world-sheet horizon and would be unphysical unless the integration constants $c_{hx}, c_{hz}$ are chosen to cancel the divergence at $\sigma_\text{ws}$. If numerators of $\vec h'(\sigma)$ vanish at the position of the world sheet horizon, then the string profile can cross $\sigma_\text{ws}$ smoothly and physical string profile solution can be obtained. 
This requirement uniquely determines the integration constants to be   
\beq
\begin{split}
c_{hx} &= \frac{2 \beta_x D_t b}{\gamma^2}, \quad c_{hz} = 2 \frac{b D_t u^3 + \beta_z D_t b}{\gamma^2}.
\end{split}
\label{chxchz}
\eeq
%Note that the requirement for the string profile to be continuous along the world sheet horizon came out naturally from the equations of motion, which is also the case when obtaining the solution for the static medium \cite{Herzog:2006gh,Gubser:2006bz} and indicates that there are no fluid gradient corrections to the position of the world-sheet horizon. With the integration constants $c_{hx}$ and $c_{hz}$ determined
The expressions \r{hprime} can then be integrated again, with the new integration
constants being fixed via the requirement that $\vec h(\sigma)$ vanishes at $\sigma\rightarrow\infty$.
Doing so yields
% and leads to rather simple form of $\vec h(\sigma)$:
\beq
\begin{split}
h_x (\sigma) =& \frac{\beta_x}{b} \left(\frac{\pi}{2} - \tan^{-1}(b\sigma)  - \frac{b \sigma}{b^2 \sigma^2 + 1}  \right) D_t b, \\
h_z(\sigma) =&  \frac{\beta_z}{b} \(\frac{\pi}{2} - \tan^{-1}(b\sigma)  - \frac{b \sigma}{b^2 \sigma^2 + 1} \) D_t b -\(\frac{\pi}{2} - \tan^{-1}(b\sigma) \) D_t u^3 \ ,
\end{split}
\label{hs}
\eeq
%In obtaining \r{hs}, the requirement of quark to be dragged with constant velocity \r{hboundary} was used to fix the second set of integration constants. 
%It turns out that solution for $\vec h(\sigma)$ is the total time derivative along the quark path of string profile $\vec x_0(\tau,\sigma)$ given in eq.~\r{x0gen} (in the fluid rest frame), in other words, 
which we can denote more simply by
\beq
b\,\vec h(\sigma) =  D_t \vec x_0(\tau,\sigma)\Big|_{\tau=0} \,,
\eeq
a result that we now see could have been guessed. So, we have shown that
%$, where $D_t$ corresponds to total time derivative.
%Therefore, in the vicinity of $\tau = 0$ the decomposition of the string profile \r{x_profile} can be done in the following way
\beq
\vec x(\tau,\sigma) = \vec x_0(\tau,\sigma) + \epsilon\, \tau D_t \vec x_0(\tau,\sigma)\Bigl|_{\tau=0} 
+ \epsilon\, b\, \vec g(\sigma) 
\label{x_profile2}
\eeq
to the order at which we are working, and our task now is to find $\vec g(\sigma)$.

%With $\vec h(\sigma)$ determined, we see that the information about the (unknown) fluid gradient corrections to the dragged string resides in the function $\vec g(\sigma)$ and we are now in position to find it.

Upon using the solution for $\vec h(\sigma)$, the Euler-Lagrange equations become 
differential equations for 
$\vec g(\sigma)$. As in the determination of $\vec h(\sigma)$, we integrate the
differential equations for $\vec g(\sigma)$ once, obtaining expressions for $\vec g'(\sigma)$.
Again as before, these expressions have poles at $\sigma=\sigma_{\rm ws}$ and the
requirement that the string profile must be regular there can be used to fix the
integration constants in the expressions for $\vec g'(\sigma)$.   Upon so doing, we find
%can be solved. Just like for the set of differential equations used to determine $\vec h(\sigma)$, the equations of motion for $\vec g(\sigma)$ involve only derivative terms, are solved to determine $\vec g'(\sigma)$ up to the integration constants. In the parallel way as for $\vec h(\sigma)$ these functions involve first order pole at $\sigma = \sigma_{\rm ws}$ unless integration constants are chosen such that the functions $\vec h(\sigma)$ are continuos when crossing the world-sheet horizon. We find the integration constants in the same way as when solving for $\vec h'(\sigma)$, and we find that $\vec g'(\sigma)$ take the following form
\beq
\begin{split}
&g_x'(\sigma) =  \frac{b^2 \beta_x \beta_z \left( \(-\pi/2 +  \tan ^{-1}(b \sigma ) \) \left(b^2 \sigma ^2+3\right) + b \sigma \right)}{ \left(b^2 \sigma ^2+1\right)^2} \partial_t u^3 +  \frac{b^2 \beta_x}{3} \partial_z u^3 \times \\
& \( \left(\gamma ^2 \left(3 \beta_z^2+1\right)+1\right)  \frac{c_1\(-b^2\sigma^2\) - c_1(-\gamma) }{\gamma ^2-b^4 \sigma ^4} - 	\frac{1}{(\sqrt \gamma + b \sigma) (\gamma + b^2 \sigma^2)} - \frac{ \frac{\pi}{2} - \tan^{-1} (b \sigma) }{1 + b^2 \sigma^2}      \) ,
\\
&g_z'(\sigma) =  b^2 \( \frac{ b^2 \sigma ^2 \left(\beta_z^2-1\right)+3 \beta_z^2-1}{\left(b^2 \sigma ^2+1\right)^2} \tan^{-1}(b\sigma) \right. \\
& \left. - \frac{1}{2} \left(\frac{\sqrt{\gamma }-b \sigma }{\gamma(b^2\sigma ^2+\gamma)} + \frac{2 \beta_z^2 (\pi -b \sigma )}{\left(b^2 \sigma ^2+1\right)^2} - \frac{\pi(1 - \beta_z^2)}{b^2 \sigma ^2+1} + \frac{1}{\gamma(b\sigma + \sqrt{\gamma})}\right)	\) \partial_t u^3 \\
& +  \frac{b^2 \beta_z \left(5 \left(b \sigma +\sqrt{\gamma }\right) \left(b^2 \sigma ^2+\gamma \right) \left(\frac{\pi }{2}-\tan ^{-1}(b \sigma )\right)-\left(b^2 \sigma ^2+1\right)\right)}{3 \left(b^2 \sigma ^2+1\right) \left(b \sigma +\sqrt{\gamma }\right) \left(b^2 \sigma ^2+\gamma \right)} \partial_z u^3,
\end{split}
\label{gs}
\eeq
where %in the $g_x'(\sigma)$ solution 
we have defined the function
\beq
\begin{split}
c_1(x) &\equiv \pi/2 - \tan^{-1}(\sqrt{-x}) - F(\sqrt{-x}) \\
&=\frac{1}{4}\(2\tan^{-1} \(\frac{1}{\sqrt{-x}}\) - \log\( \frac{(1-x)(1+\sqrt{-x})^2}{x^2} \) \).
\end{split}
\label{c1def}
\eeq
%In eq.~\r{gs} 
(The way we have chosen the signs in this definition will prove convenient later.)
We can see explicitly in (\ref{gs}) that $\vec g(\tau,\sigma)$ is regular at $\sigma=\sigma_{\rm ws}$.
It is then
possible to integrate the expressions \r{gs} analytically, fixing the integration constants
by the requirement that $\vec g(\tau,\sigma)=0$ at $\sigma\rightarrow\infty$.  The resulting
expressions for $g_x(\sigma)$ and $g_z(\sigma)$ are unwieldy and we shall not
quote them here.  In Section \ref{sec:applications} we shall, however, plot the string profile for several
choices of fluid flow and $\vec\beta$.  
%to find the actual shape of the string $\vec g(\sigma)$, but expressions are non-trivial -- the new set of integration constants would be determined by requiring that $\vec g(\sigma \to \infty) = 0$. 
In addition to being unwieldy, the expressions for $g_x(\sigma)$ and $g_z(\sigma)$ are not
of direct utility because, as we shall see in Section \ref{sec:forces}, it is 
only $\vec g'(\sigma)$ that enters into the calculation of the canonical momentum
fluxes along the string and hence of the drag force.

%, and not $g(\sigma)$ itself, enter into the calculation of canonical momentum fluxes along the string as well as drag force at the boundary, so it is not necessary to know the integrated expressions to calculate the drag force.

\section{Computing the drag force on the heavy quark}
\label{sec:forces}

In this Section we 
%define the drag force of the heavy quark which is equivalent to instantaneous quark energy and momentum loss. 
calculate the drag force acting on the heavy quark moving through the strongly coupled
fluid.  If the fluid were static, as in the original calculations~\cite{Herzog:2006gh,Gubser:2006bz},
the drag force would be a function of the temperature and the velocity $\vec \beta$ of the
heavy quark.  
%In case when the fluid is static, the drag force is a function of the fluid temperature and quark drag velocity $\vec \beta$. 
In the case that we are analyzing, where the fluid is moving but we work in the
instantaneous fluid rest-frame, the drag force again depends on $b$ and $\vec \beta$ but,
we shall show, it also depends upon the spatial gradients and time derivatives of $b$ and the
fluid $4$-velocity $u^\mu$.   After computing the drag force in the instantaneous fluid rest frame
in Section \ref{sec:forces_RF} for the case in which the fluid motion is
only along the $z$-direction, in Section \ref{sec:forces_gen} we boost the result to a frame in which the fluid at the location of
the heavy quark has some nonzero velocity in the $z$-direction, $u^3 \neq 0$.  Then, in Section \ref{sec:forcegen}
we generalize our result to the case in which the motion of the fluid is not restricted to the $z$-direction
and, in particular, may feature nonzero vorticity.

When the heavy quark is dragged through the fluid, 
in the dual gravitational description 
momentum and energy flow ``down'' the string that ``hangs down'' from the heavy quark at
$r=\infty$, trailing into the bulk metric.
In order to conserve energy and momentum, an external force must be exerted
upon the heavy quark to keep it moving at constant velocity and (in the dual
picture) to replace the energy and momentum flowing down the string.
%associated with the quark and due to energy-momentum conservation, external force has to be applied to keep the the quark moving at constant speed $\vec \beta$. 
%The presence of the external force translates to non-conservation of the stress-energy tensor
Consequently, 
\beq
\partial_\nu T^{\nu \mu} = - f^\mu(\tau) \delta^3 (\vec x - \vec \beta \tau)
\eeq
where $f^\mu(\tau)$ is the drag force acting on the heavy quark, {\it i.e.}, on the endpoint
of the string at the $r=\infty$ boundary.
The drag force at the boundary is given by~\cite{Herzog:2006gh,Gubser:2006bz}
\beq
f^\mu(\tau) = \lim_{\sigma \to \infty} n_M \int d^3 x \sqrt{-g} \mathcal T^{M\mu}
\label{force_tensor}
\eeq
where $\mathcal T^{MN}$ is the stress-energy tensor of the string obtained by varying the Nambu-Goto action \r{SNG} with respect to the $G_{MN}$, and $n_M$ is the unit-vector normal to the boundary at $r \to \infty$.  
%In the static gauge \r{static_gauge}, normal vector is pointing anti-parallel to the radial direction $n_M = -\delta_{M5}$, and relevant component of the stress-energy tensor is equal to
Because we are using the simple parametrization (\ref{static_gauge}) of the world-sheet,
the normal is simply $n_M=-\delta_{M5}$ and the relevant component of the string
stress-energy tensor is
\beq
\mathcal T^{5\mu}(\tau,\vec x,\vec y, \sigma) = \eta^{\mu\nu} \frac{1}{\sqrt{-g}} \pi^\sigma_\nu(\tau,\sigma) \delta^3(\vec y - \vec x)\,,
\label{T5mu}
\eeq
where the canonical energy/momentum fluxes along the string are obtained by varying $S_{\rm NG}$ with 
respect to $\partial_\sigma X^\mu$:
\beq
\pi^\sigma_\mu \equiv \frac{\delta S_{\rm NG}}{\delta (\partial_\sigma X^\mu)}=
- \frac{\sqrt \lambda}{2\pi} G_{\mu N} \frac{1}{\sqrt{-g}} \[ g_{\tau\sigma} \partial_\tau X^N - g_{\tau\tau} \partial_\sigma X^N \].
\label{pidef}
\eeq
Combining \r{force_tensor} and \r{T5mu}, the force acting on the quark at the boundary is given by
\beq
f^\mu(\tau) = -\frac{dp^\mu}{dt}(\tau) = -\lim_{\sigma \to \infty} \eta^{\mu \nu} \pi^\sigma_\nu(\tau,\sigma) ,
\label{force_tensor_2}
\eeq
evaluated at the location of the heavy quark, $\vec x = \vec \beta \tau$. 
Because we have used the world-sheet parameterization (\ref{static_gauge}) we
have obtained the same expression 
obtained in Refs.~\cite{Herzog:2006gh,Gubser:2006bz}; the calculation of
Ref.~\cite{Chesler:2013cqa} was done with a different world-sheet parametrization, one
for which (\ref{force_tensor})  yields an expression that differs from (\ref{force_tensor_2}). 
Note also that we are using a sign convention opposite to that in Ref.~\cite{Chesler:2013cqa}.
In the present paper, $f^\mu$ is the force exerted on the heavy quark by some external
agency (eg. an electric field) in order to keep the quark moving with constant velocity.
In the classic case of a quark moving with $\beta_z>0$ through a static plasma,
$f^z>0$ and $dp^z/dt <0$.  Note that $dp^\mu/dt$ refers to the energy/momentum
lost by the quark (lost by the quark and gained by the plasma; in the dual description,
lost by the quark and flowing down the string).

As is the case for any force,  $f^\mu(\tau)$ is not a Lorentz $4$-vector.  This is most
easily seen via the expression $f^\mu=-dp^\mu/dt$, in which $p^\mu$ is a $4$-vector
but $t$ is not boost-invariant.  We see immediately that we can define a so-called
proper force $F^\mu(\tau)$ that {\it is} a $4$-vector via
\beq
F^\mu(\tau) \equiv f^\mu(\tau) \frac{dt}{d\tau_p} = -\frac{dp^\mu}{d\tau_p} 
\label{Fprop}
\eeq
where $\tau_p$ is the (boost-invariant) proper time of the quark.
%, which does transform covariantly under Lorentz transformations. 
Because the heavy quark is moving with a constant velocity, $dt=\gamma d\tau_p$,
with $\gamma\equiv 1/\sqrt{1- |\vec\beta|^2}$ the Lorentz factor for the heavy quark.
%Since the quark is being dragged with constant velocity, the actual time and proper time can be related by constant boost, $dt = \gamma d\tau_p$. 
Then the actual drag force $f^\mu$ and the proper drag force $F^\mu$  are simply related by
\beq
f^\mu(\tau) = \frac{1}{\gamma} F^\mu (\tau).
\label{ProperVsActual}
\eeq
The distinction between actual and proper forces will play an important role in 
Sections \ref{sec:forces_gen} and \ref{sec:forcegen}.
%Section \ref{sec:forcegen} when we generalize our calculation to fluid with any values of fluid velocities and gradients. 

Just as we did in the calculation of the string profile, we 
expand the drag force in powers of the fluid gradients, writing it as
\beq
f^\mu(\tau) = f_{(0)}^\mu(\tau) + \epsilon f_{(1)}^\mu(\tau),
\eeq
where the first component $f_{(0)}^\mu(\tau)$ is the drag force when fluid gradients are neglected, first obtained in 
Refs.~\cite{Herzog:2006gh,Gubser:2006bz}, and the second component $f_{(1)}^\mu(\tau)$ is proportional to fluid gradients and is the term that we will calculate in the remainder of this Section. 
In the instantaneous fluid rest frame, in which $u^\mu = (1,0,0,0)$, 
the spatial components of the force  are given by~\cite{Herzog:2006gh,Gubser:2006bz}
\beq
\vec f_{(0), \rm RF}(\tau) = \frac{\sqrt{\lambda}}{2\pi} \frac{\gamma}{b^2} \vec \beta
\label{force_zeroth_rf}
\eeq
which shows that %for low quark velocity 
this contribution to the  force is proportional to $\gamma\vec \beta$, which is to say 
proportional to $\vec p/M$.
It is because the force is proportional to the momentum that it is referred to as a drag force.
We can then boost this result to any other frame, in which 
the fluid at the location of the heavy quark has an instantaneous
three-velocity $\vec v$ and a Lorentz factor $\gamma_v\equiv 1/\sqrt{1-|\vec v|^2}$
and, hence, 
\beq
u^\mu = \gamma_v (1,\vec v) \ .
\eeq
It is also convenient to define the $4-$velocity of the heavy quark 
\beq
w^\mu=\gamma(1,\vec \beta)\ .
\eeq
Upon boosting (\ref{force_zeroth_rf}) to a frame in which $\vec v \neq 0$, the zeroth contribution to
the drag force (i.e. the drag force obtained upon neglecting
the effects of gradients) takes the form
\beq
f_{(0)}^\mu(\tau) = -\frac{\sqrt{\lambda}}{2\pi} \frac{1}{\gamma b^2} \( s\, w^\mu + u^\mu \) ,
\label{f0gen}
\eeq
where the scalar factor $s$ is defined by
\beq
s\equiv u^\nu w_\nu\ .
\eeq
If the only nonzero component of $\vec v$ is $v_z$, we find $s=-\gamma\gamma_v(1-v_z\beta_z)$.
We shall calculate $f^\mu_{(1)}$ in the instantaneous fluid rest frame in Section \ref{sec:forces_RF}, and
in a more general frame in Section \ref{sec:forces_gen}.

Before turning to our calculation, one further general remark will prove
useful.  Starting from (\ref{force_tensor_2}) and (\ref{pidef}), it is possible to show by explicit
calculation that $w_\mu f^\mu(\tau) = 0$. Written explicitly, this takes the form
%Since the force is perpendicular to the drag $4-$velocity, $w_\mu f^\mu(\tau) = 0$, the energy loss rate is related to momentum loss rate by
\beq
\frac{dE}{dt} = \vec \beta \frac{d \vec p}{dt},
\label{onshell}
\eeq
relating the rate of energy loss to the rate of momentum loss. Since $\vec \beta = \vec p / E$ this
implies that $E^2=\vec p^2 + M^2$ for some constant $M$, which is
to say that if the quark starts out on-shell it stays on-shell.

\subsection{Drag force in the instantaneous fluid rest frame}
\label{sec:forces_RF}

We now calculate the 
canonical momentum flux along the string to first order in gradients, $\pi^{\sigma}_{\mu,(1)}(\tau,\sigma)$, and use it to obtain the corresponding drag force $f_{(1)}^\mu(\tau)$ exerted on the heavy quark
at the boundary. We calculate the drag force in the instantaneous 
fluid rest frame using the string profile given in Eqs.~(\ref{hs}) and (\ref{gs}). 
%e generalize the calculation to any frame in Section \ref{sec:forcegen}.
%As outlined previously, canonical energy/momentum fluxes 
We need to evaluate
\r{pidef} 
%are obtained by differentiating Nambu-Goto action \r{SNG} with respect to $\partial_\sigma X^\mu$,
%\beq
%\pi^\sigma_\mu(\tau,\sigma) = - \frac{\sqrt \lambda}{2\pi} G_{\mu N} \frac{1}{\sqrt{-g}} \[ g_{\tau\sigma} \partial_\tau X^N - %g_{\tau\tau} \partial_\sigma X^N \].
%\label{pi_def}
%\eeq
to linear order in $\epsilon$ after expanding
the metric $G_{\mu N}$, the induced metric $g_{ab}$, and derivatives of the string profile $\partial_a X^N$ 
in powers of $\epsilon$. % and terms of $\mathcal O(\epsilon^2)$ are neglected.
Just as for the decomposition of the string profile in Eq.~\r{x_profile2}, 
%when considering the patch located at $x^\mu = 0$ the corrections of the canonical momentum fluxes proportional to fluid gradients are found to be decomposed into two pieces
we find that
\begin{align}
\pi^\sigma_\mu(\tau,\sigma) &= \pi^{\sigma}_{\mu,(0)}(\tau) + \epsilon \( \tau \,D_t \pi^\sigma_{\mu,(0)}(\tau)\Big|_{\tau=0} + \pi^{\sigma}_{\mu,(1)}(\sigma) \).
\label{pi_decom}
\end{align}
The leading term is independent of the radial coordinate $\sigma$ and, in
the instantaneous fluid rest-frame, is given by
\beq
\pi^\sigma_{\mu, (0)}(\tau) = -\frac{\sqrt \lambda}{2\pi} \frac{1}{\gamma b^2} \(\gamma w_\mu + \delta^0_\mu \),
\label{force_firstpart}
\eeq
from which we obtain the result for the drag force absent
any effects of the fluid gradients that we already quoted in Eq.~(\ref{f0gen}).
%
%equal to minus the force given in eq.~\r{f0gen}
%\beq
%\pi^\sigma_{\mu, (0)} = -\eta_{\mu\nu} f^\nu_{(0),\rm RF},
%\label{pi0}
%\eeq 
%
The term proportional to time $\tau$ in (\ref{pi_decom}) is given by
%is obtained by taking total time derivative of \r{pi0} and setting $u^z = 0$ in the end,
\beq
\begin{split}
&  D_t \pi^\sigma_{\mu,(0)}(\tau)\Big|_{\tau=0} = \frac{\sqrt \lambda}{2\pi} \frac{1}{\gamma b^2} \times \[ \frac{2}{b} (\gamma w^\mu - \delta^\mu_0) D_t b + (w^\mu w^z + \delta^\mu_3) D_t u^z \],
\end{split}
\label{force_secondpart}
\eeq
where 
$D_t$ was defined in (\ref{convective_derivative}).
We can neglect this term since it appears in (\ref{pi_decom}) multiplied
by $\tau$ and we are evaluating the drag force on the heavy quark at $\tau=0$.
%$D_t b = \partial_t b + \partial_i b$ and analogously for $u^z$. 
%Therefore,  the first 2 terms in \ref{pi_decom} are found from the canonical flux at leading order and do not carry no new information. 

%The actual correction of the drag force due to gradients are encoded in 
The nontrivial part of the computation is the determination of
$\pi^\sigma_{(1)}(\sigma)$. After collecting terms proportional to $\epsilon$ at $\tau = 0$, we find that 
%canonical momentum fluxes due to fluid gradients are given by
\beq
\begin{split}
\pi^\sigma_{x, (1)}( \sigma) &= \frac{\sqrt \lambda}{2\pi b} \gamma \beta_x \[\beta_z \left(\frac{b \sigma}{b^2 \sigma ^2+1} + \pi-2 \tan ^{-1}(b \sigma ) -\sqrt{\gamma }\right) \partial_t u^3 + \right. \\ 
& \left. \frac{ \left(\frac{ b \sigma }{b^2 \sigma ^2+1} - \sqrt{\gamma }+ \left(1 + 3 \beta_z^2\right) \left(\frac{\pi}{2} - \tan ^{-1}(b \sigma ) \right ) - c_1\left(-\gamma \right) \left(\gamma ^2 \left(1 - 3 \beta_z^2\right) + 1\right) \right)}{3} \partial_z u^3 \right ], \\
\pi^\sigma_{z, (1)}(\sigma) &= \frac{\sqrt{\lambda}}{2\pi b} \gamma \[  \left(-\frac{\beta_x^2  b\sigma}{b^2 \sigma ^2+1} + \left(2 \beta_z^2+1\right)\left(\frac{\pi}{2} - \tan ^{-1}(b \sigma ) \right) - \sqrt{\gamma} \left(1-\beta_x^2\right)\right) \partial_t u^3  + \right. \\
& \left.  \frac{\beta_z \left(\frac{b\sigma }{b^2 \sigma ^2+1} + (1 - 3\beta_x^2)
\left( \frac{\pi}{2} - \tan ^{-1}(b \sigma ) \right) - \sqrt{\gamma} - c_1\left(-\gamma\right) \left(\gamma ^2 \left(1 - 3 \beta_z^2\right) - 5\right)
\right)}{3} \partial_z u^3  \right ],
\end{split}
\label{pi1}
\eeq
where $c_1(-\gamma)$ was defined in \r{c1def} and the conservation of the stress-energy tensor \r{eos} has 
been used to eliminate $\partial_t b$ and $\partial_z b$ in favor of $\partial_t u^3$ and $\partial_z u^3$.
We now determine the contributions of these 
canonical momentum fluxes to the drag force on the heavy
quark at the boundary, which is to say
we take the $\sigma\to\infty$ limit.
%Now canonical momentum fluxes can can be calculated at the boundary $\sigma \to \infty$ to find the drag force for infinitely massive quark. 
The terms $\frac{b \sigma}{b^2 \sigma^2 + 1}$ and $\frac{\pi}{2} - \tan^{-1}(b\sigma)$ vanish in
this limit, and the contribution to the drag force that is first order in gradients is given by
\beq
\begin{split}
f^x_{(1)} &= \frac{\sqrt \lambda}{2\pi b} \gamma \beta_x \(  \sqrt \gamma \beta_z \partial_t u^3 + 
 \frac{ \sqrt{\gamma} + c_1\left(-\gamma\right) \left(\gamma ^2 \left(1 - 3 \beta_z^2\right) + 1\right)    }{3} \partial_z u^3 \),  \\
f^z_{(1)} &= \frac{\sqrt \lambda}{2\pi b}\gamma \( \sqrt{\gamma}(1-\beta_x^2) \partial_t u^3
+  \frac{\sqrt{\gamma}+ c_1\left(-\gamma\right) \left(\gamma ^2 \left(1 - 3 \beta_z^2\right) - 5\right) }{3} \beta_z \partial_z u^3 \),
\end{split}
\label{forcesRF}
\eeq
with the $t$ component of the force given by $f^t_{(1)} = \beta_x f^x_{(1)} + \beta_z f^z_{(1)}$, ensuring that the quark stays on shell. 
The complete expression for the drag force is obtained by combining the contributions
from (\ref{force_firstpart}) and %(\ref{force_secondpart}) with 
(\ref{forcesRF}).

Before turning to generalizations of this result, we end this subsection by remarking that
%First, note that 
both the terms proportional to $\partial_t u^3$ and the terms
proportional to  $\partial_z u^3$ in (\ref{forcesRF}) are proportional 
to $\gamma^{3/2}$ for large $\gamma$.  This is apparent
for the terms proportional to $\partial_t u^3$.  To see this for
the terms proportional to $\partial_z u^3$, note that
in the large-$\gamma$ limit
\beq
c_1(-\gamma) = -\frac{1}{3 \gamma^{3/2}} + \mathcal O\(\frac{1}{\gamma^{2}}\)\ .
\label{c1_asym}
\eeq
This means that for large enough $\gamma$, the contributions
to the drag force that are first order in fluid gradients, namely (\ref{forcesRF}),
dominate over the zeroth order expression (\ref{force_zeroth_rf}) for the drag force
in the absence of fluid gradients.   Comparing (\ref{force_zeroth_rf}) and (\ref{forcesRF})
we see that the first order contributions to the drag force are smaller than the zeroth order contributions
when
\beq
\sqrt{\gamma}<\frac{1}{ b \,\left|\partial_t u^3\right|}  \quad {\rm and}\quad \sqrt{\gamma} < \frac{9}{5} \frac{1}{b\, \left|\partial_z u^3 \right|}
\eeq
or, using (\ref{eos}), when
\beq
\sqrt{\gamma}<\frac{1}{ \left| \partial_z b\right|} \quad {\rm and}\quad \sqrt{\gamma} < \frac{3}{5}\frac{1}{\left| \partial_t b\right|}\ ,
\label{limit_gamma}
\eeq
with $\gamma$ and the gradients on the right-hand sides of all these expressions evaluated in the
frame of reference in which the fluid is instantaneously at rest at the location of the moving heavy quark.
This result suggests that at larger values of $\gamma$
the expansion of the drag force in powers of the fluid gradients may break down, although
to be sure of this it would be useful to extend our calculation to higher order in gradients.
At a qualitative level, what seems to be happening is that at large enough $\gamma$ the 
heavy quark sees a gradient in the fluid as sudden, and the gradient expansion of
the drag force ceases to be valid. 
Note that the criterion for the validity of the hydrodynamic description
of the fluid itself is $|\partial_z b| \ll 1$ and $|\partial_t b|\ll1$, meaning that
as long as the motion of the fluid is described well by hydrodynamics the 
limitation (\ref{limit_gamma}) on the values of $\gamma$ at which the
gradient expansion can be used to describe the drag force on the heavy quark
sets in at some $\gamma \gg 1$.  As hydrodynamics itself breaks down,
the range of validity of the gradient expansion in the calculation of the 
heavy quark drag force becomes smaller and smaller.

Note that for quarks with finite $M$ the description
of the drag force in terms of a single trailing string is only valid 
for~\cite{Liu:2006he,CasalderreySolana:2007qw,CasalderreySolana:2011us} 
\beq
\sqrt{\gamma}\ll \frac{M}{T\sqrt{\lambda}}\ ,
\label{old_limit_gamma}
\eeq
since the external force required to move a quark with mass $M$ at a larger $\gamma$
would result in copious pair-production of quark-antiquark pairs.
However,  we are working
in the $M\rightarrow\infty$ limit throughout this paper, meaning that
the criterion (\ref{old_limit_gamma}) by itself would allow us to consider arbitrarily large $\gamma$.
Instead, even in the $M\rightarrow\infty$ limit
the magnitude of the fluid gradients imposes new, lower, limits (\ref{limit_gamma}) on how large $\gamma$ can be,
at least if one wishes to use a gradient expansion to calculate the drag force.  These considerations
motivate future extensions of our calculations, both to higher order in fluid gradients and
to finite quark mass $M$.  An analysis in which one takes the $\gamma\rightarrow\infty$ limit
first, with finite mass quarks, and only later takes $M\rightarrow\infty$ would necessarily look
very different from the analysis in this paper.

\subsection{Generalizing to a frame in which the fluid is moving}
\label{sec:forces_gen}

In Section \ref{sec:forces_RF} we have calculated the drag force exerted on a heavy quark
moving through the fluid, in the instantaneous fluid rest frame and in a fluid
that is moving only along the $z$-direction, obtaining the result (\ref{forcesRF}).
We can now boost this result to a frame in which the fluid at the location
of the heavy quark has velocity $\vec v=(0,0,v_z)$, instead of being at rest.
We do this by first constructing the proper force $F^\mu$ from $f^\mu$, according
to (\ref{ProperVsActual}), then applying a Lorentz transformation to the $4-$vector $F^\mu$, bringing
it to the desired frame, then working out the value of $\gamma$ in the desired frame,
and finally using (\ref{ProperVsActual}) again to obtain $f^\mu$ in the new frame.
The calculation, which is tedious but straightforward, yields the following expression
for the drag force exerted on a heavy quark moving with velocity $\vec \beta$ 
through a fluid that is moving only along the $z$-direction and that has  velocity $\vec v=(0,0,v_z)$
at the location of the heavy quark:
\beq
\begin{split}
f^x_{(1)} &= -\frac{\sqrt \lambda}{2\pi}\frac{s \gamma_v^2 \beta_x }{3b} \Bigl[  
\partial_t v_z \[ c_1(s) \left(s^2+1\right)  \gamma_v v_z+3 c_1(s)   s ( s \gamma_v + \gamma) \Delta \beta_z + \sqrt{-s} \gamma_v (3 \Delta \beta_z +v_z) \]  \Bigr. \\
&\Bigl. + \partial_z v_z \left[c_1(s)  \left(s^2+1\right) \gamma_v + 3 c_1(s)  s (s \gamma_v v_z + \gamma \beta_z )  \Delta \beta_z +  \sqrt{-s} \gamma_v (3 v_z \Delta \beta_z + 1)\right] \Bigr],
\end{split}
\label{forcesNotRF1}
\eeq
\beq
\begin{split}
f^z_{(1)} &= -\frac{\sqrt \lambda}{2\pi} \frac{\gamma_v^2}{3b\gamma} \Bigl[ \partial_t v_z \Bigl(c_1(s) \gamma_v \left[\gamma_v \left(s^2 \left(v_z^2-3\right)+v_z^2\right)+s\gamma \left(\left(s^2+1\right)\beta_z v_z-3\right)  \right] \Bigr. \Bigr.  \\
&\Bigl. \Bigl. + 3   c_1(s) s \gamma \Delta \beta_z \left[ s \beta_z (s \gamma_v +\gamma)-\gamma_v v_z \right]  +\gamma_v \sqrt{-s}  \left[ s \gamma \beta_z  (3 \Delta \beta_z +v_z)+\gamma_v \left(v_z^2-3\right)\right] \Bigr) \Bigr. \\
&\Bigl. + \partial_z v_z \Bigl(3 c_1(s) s \gamma \beta_z \Delta \beta_z  \left[\left(s^2-1\right) \gamma_v v_z + s \gamma \beta_z  \right]+c_1(s) \gamma_v \left[s^3 \gamma \beta_z   - 2 s^2 \gamma_v  v_z  \right. \Bigr. \Bigr. \\
&\Bigl. \Bigl. \left.+ s \gamma (3 v_z-5 \beta_z) +\gamma_v v_z \right]  + \sqrt{-s} \gamma_v  \left[s \gamma \beta_z (3 v_z \Delta \beta_z + 1)-2 \gamma_v v_z\right]\Bigr) \Bigr].
\end{split}
\label{forcesNotRF}
\eeq
Here,
$\Delta \beta_z$ denotes the (relativistic) difference between the velocities of the quark  and the
 fluid in the $z-$direction
\beq
\Delta \beta_z =  \frac{\beta_z - v_z}{1- \beta_z v_z} = \frac{\gamma \gamma_v}{s} (v_z - \beta_z).
\eeq 
Recall that our notation is such that $\vec v$ is the velocity of the fluid, here in the $z$-direction,
$\gamma_v = 1/\sqrt{1-v_z^2}$ is the fluid velocity Lorentz factor, and
$u^\mu = \gamma_v(1, \vec v)$. Furthermore, $\vec \beta$ is the velocity
of the heavy quark, $\gamma=1/\sqrt{1-\vec\beta^2}$, $w^\mu=\gamma(1,\vec\beta)$, and the scalar factor $s$ is given by
\beq
s \equiv u^\mu w_\mu = -\gamma \gamma_v (1 - v_z \beta_z)\ .
\eeq
(Note that in the instantaneous fluid rest frame $s=-\gamma$. We chose the signs in our definition
(\ref{c1def}) of the function $c_1$ such that henceforth what will appear in many equations is $c_1(s)$.)
In the next subsection, we shall find a much more compact way of writing the result
(\ref{forcesNotRF}) after first generalizing our calculation of the drag force to the
case in which the fluid can move in any direction.

\subsection{General fluid motion}
\label{sec:forcegen}

Although in the explicit applications of our results that
we shall present in Section 4 we shall only need the results
we have already obtained in Sections \ref{sec:forces_RF} and \ref{sec:forces_gen}, before proceeding
we now wish to generalize our analysis beyond the case in which the
motion of the fluid is only along a single axis to consider any possible
three-dimensional motion of the fluid satisfying the hydrodynamic equations of 
motion (\ref{HydrodynamicEquations}).  
It will turn out that generalizing our analysis in this
way will yield a more compact form of our result that
is more user-friendly than (\ref{forcesNotRF1}) and (\ref{forcesNotRF}),
in addition to being more general.
We will continue to work only
to first order in fluid gradients, but we will no longer restrict to the case
(\ref{VanishingGradients}). That is, we will allow all the 
velocity gradients and time derivatives $\partial_\alpha u_\beta$ to be nonzero, but will
continue to assume that they are small enough in magnitude that second
and higher derivatives can be neglected.  The time derivative and
gradients of the temperature are then determined from 
$\partial_\alpha u_\beta$ via the hydrodynamic equations in the form
(\ref{partialb}).
We will start by  writing down
the most general general possible Lorentz covariant proper drag
force $F^\mu$, related to $f^\mu$ by (\ref{ProperVsActual}), to first
order in $\partial_\alpha u_\beta$, and will then use the calculations that
we have done already (plus a little bit more)  to fix all the coefficients
in the general expression. In this way we will obtain the drag force $f^\mu$
up to first order in $\partial_\alpha u_\beta$ for a general fluid configuration.

To zeroth order in gradients, we already have the general result
for $f^\mu_{(0)}$ in (\ref{f0gen}), in explicit form.   We now write
a general, but formal, expression for the contribution to 
the drag force $f_{(1)}^\mu$ that is first order in the fluid gradients
$\partial_\alpha u_\beta$ by writing the most general possible
Lorentz covariant vector $F^\mu_{(1)}$ and, from (\ref{ProperVsActual}),
dividing by $\gamma$:
%
%In this Appendix, the construction of the general covariant form of the force is discussed.
%From eq.~\r{Fprop} and the fact that $f^\mu_{(1)}$ is proportional to $\partial_\alpha u_\beta$, the most general expression for the gradient correction to force can be written as
%
\beq
\begin{split}
f^\mu_{(1)} &= \frac{\sqrt \lambda}{2\pi}\frac{1}{b\gamma} \[ a_1  \eta^{\mu \alpha}  w^\beta + u^\mu (a_2 \eta^{\alpha \beta} + a_3 w^\alpha w^\beta + a_4 u^\alpha w^\beta ) \right. \\
& \left.+ w^\mu ( a_5  \eta^{\alpha \beta} + a_6  w^\alpha w^\beta + a_7 u^\alpha w^\beta )  + \eta^{\mu \beta} (a_8 u^\alpha + a_9 w^\alpha ) \right.\\
&+\left. \epsilon^{\lambda\nu\alpha\beta} \( a_{10} \eta^{\mu}_{~\lambda} u_\nu + a_{11} u^\mu w_\lambda u_\nu + a_{12} w^\mu w_\lambda u_\nu + a_{13} \eta^{\mu}_{~\lambda} w_\nu \)  \] \partial_\alpha u_\beta
\end{split}
\label{f_general_aterms}
\eeq
where the coefficients $a_{1}\ldots a_{13}$ are (at present arbitrary) functions of the only possible 
Lorentz scalar that is zeroth order in derivatives, namely  the now
familiar $s\equiv u^\alpha w_\alpha$.  Our task now is to determine $a_{1}\ldots a_{13}$.
The terms multiplied by $a_1\ldots a_9$ in (\ref{f_general_aterms}) are all the possible terms that can be written
without introducing $\epsilon^{\lambda\nu\alpha\beta}$. This can be seen by
noting that the index $\mu$ can be placed on the gradient direction $\partial^\mu$ (the $a_1$ term), on 
the fluid velocity $4-$vector $u^\mu$ (the $a_2\ldots a_4$ terms), on the heavy quark
velocity $4-$vector $w^\mu$ (the $a_5\ldots a_7$ terms), or on the fluid velocity $4-$vector 
that is acted upon by the derivative (the $a_8$ and $a_9$ terms).  
(We have used $u^\beta \partial_\alpha  u_\beta = 0$, 
a consequence of $u^\beta u_\beta=-1$, to eliminate other terms.)
The terms multiplied by $a_{10}\ldots a_{13}$
are the only allowed terms that can be constructed by
contracting with the totally antisymmetric tensor $\epsilon^{\lambda\nu\alpha\beta}$.
%Note that due to the normalization of the fluid $4-$velocity, which ensures that $u^\alpha \partial_\beta u_\alpha = 0$, the number of possible terms is reduced.
For example $a_{10}$ multiplies the fluid vorticity $\tilde\omega^\mu$, defined in (\ref{vorticity_definition}).
Note, however, that there is a sense in which effects of vorticity are hiding among the $a_1\ldots a_9$ terms
because since
 \beq
\epsilon^{\mu\nu\alpha\beta} \tilde \omega_\nu w_\alpha u_\beta = \frac{1}{2}\[ \( \eta^{\mu \beta} w^\alpha - \eta^{\mu \alpha} w^\beta \) - \( u^\mu w^\beta - \eta^{\mu \beta} s \) u^\alpha  \]\partial_\alpha u_\beta
\label{c10}
\eeq 
there is one linear combination of the $a_1$, $a_4$, $a_8$ and $a_9$ terms
that vanishes if $\tilde\omega=0$, a fact that will be relevant. 

There is one completely general constraint on $f^\mu$ that we have not
yet imposed, namely
$w_\mu f^\mu_{(1)} = 0$. 
Using (\ref{f_general_aterms}), this constraint takes the form
\begin{align}
&\[ \eta^{\alpha \beta} ( s \,a_2 - a_5) + u^\alpha w^\beta  ( s \,a_4 + a_8 - a_7) \notag \right. \\
&\left. + w^\alpha w^\beta (a_1 + s\, a_3 - a_6 + a_9) + \epsilon^{\lambda\nu\alpha\beta} w_\lambda u_\nu \(a_{10} + s \,a_{11} - a_{12} \)
 \] \partial_\alpha u_\beta = 0
\label{f_global2}
\end{align}
and since the relation has to be satisfied for the arbitrary vectors $u^\mu$ and $w^\mu$ independently, four 
out of the 13 coefficients $a_1\ldots a_{13}$ can be eliminated, {\it e.g.}, 
\beq
%\begin{split}
a_1 = a_6 - a_9 - s\, a_3, \quad s \,a_2 =  a_5 , \quad s\, a_4 = a_7 - a_8, \quad
a_{12} = a_{10} + s\, a_{11}. 
\label{a_relations}
%\end{split}
\eeq
In this way we can replace (\ref{f_general_aterms}) by 
\beq
\begin{split}
f_{(1)}^\mu &= -\frac{\sqrt{\lambda}}{2\pi} \frac{1}{b\gamma} \[ 
c_1(s) w^\beta (u^\mu w^\alpha -s\, \eta^{\mu \alpha} ) + 
c_2(s) \eta^{\alpha \beta} (u^\mu + s \,w^\mu) + 
c_3(s) w^\beta( w^\mu w^\alpha + \eta^{\mu \alpha} ) \right. \\
&+\left. c_4(s) u^\alpha w^\beta (u^\mu + s\, w^\mu) + 
c_5(s) u^\alpha( u^\mu w^\beta - s\, \eta^{\mu \beta} ) + 
c_6(s) (w^\alpha \eta^{\mu \beta} - \eta^{\mu \alpha} w^\beta) \right. \\
&\left. + \epsilon^{\lambda\nu\alpha\beta} \( 
c_7(s) \(\eta^{\mu}_{~\lambda} + w^\mu w_\lambda \) u_\nu + 
c_8(s) \( u^\mu + s w^\mu \) w_\lambda u_\nu + 
c_9(s) \eta^{\mu}_{~\lambda} w_\nu \) 
 \] \partial_\alpha u_\beta
\end{split}
\label{f_global}
\eeq
%
%
%In this way, the general expression for the force in eq.~(\ref{f_global}) is constructed with the coefficients given by
%\beq
%\begin{split}
%c_1 &= a_3, \quad c_2 = a_5/(u \cdot w), \quad c_3 = a_6, \\
%c_4 &= a_7/ (u\cdot w), \quad c_5 = -a_8/ (u \cdot  w), \quad c_6 = a_9.
%\end{split}
%\eeq
with a new set of nine unknown coefficients $c_1\ldots c_9$ that are each still unknown
functions of $s$ that are related to the $a_1\ldots a_{13}$ by
\begin{alignat}{3}
  c_1&=a_3, && s\,c_2 = a_5, &&c_3 = a_6, \notag\\
  s\, c_4&=a_7 ,\quad &&s\, c_5 = -a_8,\quad  &&c_6 = a_9, \notag \\
  c_7 &= 2 a_{10} + s\,a_{11} , \qquad &&c_8 =s\, a_{10}+ (1+s^2)a_{11}  , \qquad &&c_9 = a_{13},
\end{alignat}
%Note that when constructing general solution, either the form (\ref{f_global}) in terms of $c_{1-9}$ or (\ref{f_global2}) in terms of $a_{1-13}$ can be used since the actual calculated drag force is perpendicular to $4-$vector $w^\mu$, but when using the expression \r{f_global} it is explicitly clear that the quark is on-shell.
with $a_1$, $a_2$, $a_4$ and $a_{12}$ related to the $c$'s through (\ref{a_relations}).
Note that the combination of terms (\ref{c10}) that vanishes if the vorticity vanishes
is now a linear combination of the terms multiplied by $c_5$ and $c_6$.

We can now attempt to use the results of our previous calculation, namely (\ref{forcesRF}), to fix
the coefficients $c_1\ldots c_9$ in (\ref{f_global}).  We start by writing (\ref{f_global}) in the
instantaneous fluid rest frame, in which $u^\mu=(1,0,0,0)$ and $s=-\gamma$.  We then
restrict to the fluid motion that we analyzed in Sections \ref{sec:string_quark} and \ref{sec:forces_RF}, which is to say that we set 
the partial derivatives (\ref{VanishingGradients}) to zero, keeping only those in (\ref{NonzeroGradients}).
We then compare the expressions for $f^x_{(1)}$ and $f^z_{(1)}$
so obtained with the expressions in 
(\ref{forcesRF}), term by term.
By ``term by term'' we mean  that we compare those terms in $f^x_{(1)}$ (or $f^z_{(1)}$)
from (\ref{f_global}) and (\ref{forcesRF}) that are proportional to $\partial_z u^3$ (or $\partial_t u^3$)
and that are proportional to $\beta_z^0$ or $\beta_z$ or $\beta_z^2$ or $\beta_z^3$.
In this sense, we make 16 comparisons between (\ref{f_global}) and (\ref{forcesRF}), resulting in
16 expressions that specify various of the $c$'s.  What we find when we do this exercise is that
we only obtain 5 independent constraints on the $c$'s, and that these constraints can
be used to fix the values of $c_1\ldots c_4$ and $c_5+c_6$.  However, we cannot
determine $c_5-c_6$ or $c_7\ldots c_9$.   This is not unexpected, since by setting
the partial derivatives  (\ref{VanishingGradients}) to zero we have set the vorticity to zero
and have ensured that the terms multiplied by $c_7\ldots c_9$ in (\ref{f_global}) all vanish
as does (\ref{c10}).  

From the above exercise we conclude that in order to complete the determination of all
the $c_1\ldots c_9$ we need to analyze a fluid configuration with nonzero vorticity.
We have repeated the analysis of Sections \ref{sec:string_quark} and \ref{sec:forces_RF} for a fluid in which 
$\partial_t u^1\neq 0$, $\partial_x u^1 \neq 0$, $\partial_x u^3\neq 0$ and $\partial_z u^1\neq 0$,
in addition to the nonzero partial derivatives in (\ref{NonzeroGradients}).  We also allowed
$\vec\beta$ to have nonzero components in all three directions.  As a check, we first considered 
the case where $\partial_x u^3 = \partial_z u^1\neq 0$, which is to say we did a much
more complicated calculation than in Sections \ref{sec:string_quark} and \ref{sec:forces_RF} but still with vanishing vorticity.
We then repeated the exercise described in the preceding paragraph and once again
found only 5 independent constraints on the $c$'s that served to fix $c_1\ldots c_4$
and $c_5+c_6$.  So, we obtained no new information at all.  We then redid all the
calculations with $\partial_x u^3 \neq \partial_z u^1$.  In this case, we found 
9 independent constraints on the $c$'s that, finally, served to fix them all.  We find:
%So far we considered configuration with only two non-zero gradients $\partial_t u^z$ and $\partial_z u^z$ and such configuration for non-zero drag velocity components in $x-$ and $z-$directions is sufficient to determine coefficients $c_{1-4}(s)$ as well as sum of coefficients $c_5(s)$ and $c_6(s)$, but the configuration is not sufficient to determine the difference $c_5(s) - c_6(s)$, neither to fix remaining coefficients $c_{7-9}(s)$ uniquely.  In order to determine coefficients $c_{5-9}(s)$ uniquely, we considered the case when quark has all three non-zero drag velocity components,  $\vec w = (w_x, w_y, w_z) \neq 0$ and fluid velocities in $x-$ and $z-$ directions depended on $t, x, z$ space-time coordinates, $u^1(t,x,z), u^3(t,x,z)$. With such configuration, we were able to determine all coefficients $c_{1-9}(s)$ in eq.~\r{f_global} and we found that  
%\footnote{
%In the fluid rest frame, $u^\mu = (1,\vec 0)$, so the product of $4-$velocities is equal to $s  = - \gamma$, and \r{f_global} can be directly compared with the analog of eq.~\r{forcesRF}.}
%\beq
%\begin{split}
%c_1 &= \pi/2 - \tan^{-1} (\sqrt{-u \cdot w}) - F( \sqrt{-u \cdot w} ), \\
%c_2 &= \frac{1}{3} \( \sqrt{-u \cdot w} + (1 + (u \cdot w)^2) c_1 \), \\
%c_3 &=  (u \cdot w) c_6 = -(u \cdot w) c_1, \\
%c_4 &= - c_5 = \frac{1}{\sqrt{-u \cdot w}} - (u \cdot w) c_1.
%\end{split}
%\eeq
\beq
\begin{split}
c_1(s) &= \pi/2 - \tan^{-1} (\sqrt{-s}) - F( \sqrt{-s} ), \\
c_2(s) &= \frac{1}{3} \( \sqrt{-s} + (1 + s^2) c_1(s) \), \\
%\notag
%\end{split}
%\eeq
%\beq
%\begin{split}
c_3(s) &=   c_6(s) = -s c_1(s), \\
c_4(s) &= - c_5(s) = \frac{1}{\sqrt{-s}} - s c_1(s), \\
c_7(s) &= c_8(s) = c_9(s) = 0,
\label{cdef}
\end{split}
\eeq
where $c_1(s)$ is the same function as defined in \r{c1def} previously.
As a nontrivial check of the calculation, we note that we obtained the same
results for $c_1\ldots c_4$ and $c_5+c_6$ when we fixed them via our calculations
for configurations with or without vorticity.  As another nontrivial check,
we have used (\ref{f_global}) with (\ref{cdef}) to reproduce our
results (\ref{forcesNotRF1}) and (\ref{forcesNotRF}) from Section \ref{sec:forces_gen}.

Although we included $c_7\ldots c_9$ for completeness, we could have argued
from the beginning that they must vanish. If any of these coefficients were nonzero,
there would be a contribution to the drag force that was proportional to the 
vorticity, or to one of the other expressions involving an explicit $\epsilon^{\mu\nu\alpha\beta}$.
This would violate time-reversal and parity symmetry.  It might be  interesting
to repeat our analysis for a (chiral) fluid in which these symmetries are in fact violated at
a microscopic level.  We expect that in such a fluid these coefficients could be nonzero.
Note, however, that $c_5\neq c_6$ in our calculation. This means that the presence of nonzero vorticity in the 
${\cal N}=4$ SYM fluid that we have analyzed {\it does} 
affect the drag force that the fluid exerts on a heavy quark moving through it, via a contribution to
the drag force that is proportional to (\ref{c10}).

The most general result of this paper is the expression (\ref{f_global}) which, with (\ref{cdef}), 
gives the contribution to the drag force on a heavy quark moving through the strongly coupled fluid in arbitrary
hydrodynamic motion that is first order in fluid gradients.  By rearranging terms we have found
a more compact version of this expression:
%Finally, by rearranging the terms in ~\r{f_global}, drag force can be expressed in the following form
%\beq
%\begin{split}
%f_{(1)}^\mu = -\frac{\sqrt \lambda}{2\pi} \frac{1}{b\gamma_w} \[ c_1(s) (u^\mu D_w s - s \partial^\mu s - s(s D_u + D_w) Q^\mu) + c_2(s) Q^\mu \partial_\alpha u^\alpha - \sqrt{-s} D_u Q^\mu \]
%\label{f_final}
%\end{split}
%\eeq
\beq
\begin{split}
f_{(1)}^\mu =& -\frac{\sqrt \lambda}{2\pi} \frac{1}{b\gamma} \times \\
& \Bigl[ c_1(s) \bigl(u^\mu w^\alpha \partial_\alpha s - s \partial^\mu s - s(s u^\alpha + w^\alpha)\partial_\alpha U^\mu\bigr) + c_2(s) U^\mu \partial_\alpha u^\alpha - \sqrt{-s} u^\alpha \partial_\alpha U^\mu \Bigr],
\label{f_final}
\end{split}
\eeq
%with total derivatives along fluid and drag velocities given by $D_u \equiv u^\alpha \partial_\alpha, D_w \equiv w^\alpha \partial_\alpha$, and 
where $U^\mu \equiv u^\mu + s w^\mu$ denotes the component of fluid $4-$velocity $u^\mu$ that is
perpendicular to the heavy quark  $4-$velocity $w^\mu$. This (deceptively) compact 
expression for the drag force arising due to fluid gradients at first order is
the main result of this paper.  The explicit results given in earlier subsections that we shall
employ in Section 4 are all special cases of (\ref{f_final}).

\section{Applications}
\label{sec:applications}

In this Section we shall apply our result (\ref{forcesRF}) 
and its generalization (\ref{f_final})
%--- a special case of our general result (\ref{f_final}) --- to
to analyze the effects of fluid gradients on the drag force on a heavy quark in three settings, ordered by
increasing complexity. We will first consider a quark at rest in the instantaneous fluid rest frame, and
show that even in this case the fluid can exert a ``drag'' force on the heavy quark.  We will then
consider two applications of our result to models of interest in the context of heavy ion collisions.
In Section \ref{sec:bj} we consider boost-invariant expansion of the fluid, \`a la Bjorken.  In Section \ref{sec:collsh} we return to the colliding sheets of energy whose analysis in Ref.~\cite{Chesler:2013cqa} provided
the initial motivation for the present study, as we have described in Section~\ref{sec:intr}.

\subsection{A quark at rest in a fluid that is, instantaneously, at rest}

%For simplicity, let's concentrate on the drag force of quark moving in the static fluid with non-vanishing gradients, eqs.~\r{forcesRF}. 

As a very simple example with which to illustrate how fluid gradients can have
nontrivial consequences for the ``drag'' force exerted by the fluid on a heavy 
quark, let us consider a heavy quark that is at rest in a fluid that is instantaneously at rest at
the location of the heavy quark.
However, the fluid is not static and is not spatially uniform.  If we neglect the effects of gradients, 
there would be no force on the quark:  the quark is not moving, the fluid is not moving, so there
can be no drag force.  For simplicity let us consider the case where the fluid motion is only
in the $z$-direction, as in Section \ref{sec:forces_RF}.  In this case, from (\ref{forcesRF}) we see that 
as a consequence of the time variation of the fluid velocity there {\it is} a force acting on the quark, pushing it
in the $z$-direction, namely
%
%When quark is at rest, $\vec \beta = 0$, the only contribution to the force arises from fluid velocity gradient along the time direction
\beq
 f^z_{(1)} = \frac{\sqrt \lambda}{2 \pi b}\partial_t u^3\,,
 \label{force_at_zero_speed}
\eeq
even though $\vec\beta=0$.  This shows that the force exerted by the fluid
on the heavy quark cannot always be thought of as  a drag force, a point that
was already made in Ref.~\cite{Chesler:2013cqa}.
Note, however, that the sign of the force (\ref{force_at_zero_speed}) 
is consistent with an interpretation in terms of drag with a time delay.
If $\partial_t u^3 >0$, then a short time ago $u^3$ was negative.
That means that if we think in terms of drag we would expect that
a short time ago the fluid was pushing the quark toward negative $z$,
which in turn means that a short time ago the external agency 
holding the quark at constant $\vec\beta=0$ 
would have been exerting a force $f^z>0$.  So, we can interpret
(\ref{force_at_zero_speed}) in terms of a time delay in the 
response of the drag force to changing fluid conditions.  
Comparing (\ref{force_at_zero_speed}) to (\ref{f0gen}), we can estimate
that the time delay is of order $b$ for a quark at rest.
The results of Ref.~\cite{Chesler:2013cqa} indicate that a time delay
like this is generic. Such a time delay has also been seen in Ref.~\cite{Guijosa:2011hf}.

\subsection{Bjorken Flow}
\label{sec:bj}

In 1982 Bjorken discovered a simple solution to the
zeroth order (ideal) hydrodynamic equations
of motion~\cite{Bjorken:1982qr} 
that has since then often been used as a toy model for the longitudinal expansion
of the fluid produced in heavy ion collisions.  In Bjorken's solution, the fluid
expands in the $z$-direction only and its expansion is boost invariant.
The fluid $4$-velocity is given by
\beq
u^\mu = \(\frac{t}{\tau_p},0,0,\frac{z}{\tau_p}\),
\label{uBF}
\eeq
where $\tau_p \equiv \sqrt{t^2 - z^2} $ is the proper time,
which is to say
\beq
\vec v = \(0,0,\frac{z}{t}\)\ .
\eeq
The solution is only defined in the forward light-cone, $z>|t|$. 
The temperature of the fluid, and hence its energy density and
pressure, depend only on  $\tau_p$.
We shall refer to this solution to hydrodynamics as Bjorken flow.
If the fluid were ideal, with no viscosity and hence no contribution
to the fluid stress-energy tensor from gradients, 
then $b(\tau_p)\propto \tau_p^{1/3}$~\cite{Bjorken:1982qr}.
This dependence is modified when nonzero viscosity
and hence effects of gradients are taken into account,
as for example in Refs.~\cite{Baier:2007ix,Chesler:2009cy}.
The gravitational dual of Bjorken flow was first constructed in Ref.~\cite{Janik:2005zt}.
For us, though, the calculation of
the drag force on a heavy quark in a fluid
expanding in a Bjorken flow is simply a special case of the
calculation we have presented in Section \ref{sec:forces_gen}.  We just need to
apply the result (\ref{forcesNotRF}) or, in its more
general form, the result (\ref{f_final}),  to the velocity profile \r{uBF}. 
The temperature could be obtained from \r{uBF} but we
will not need to do so, as we will leave our result written
in terms of $b(t,z)=b(\tau_p)$.

%, but for our analysis, because gravitational dual for Bjorken Flow is just a special case of the drag force calculation that we carried out, it is enough to know velocity profile \r{uBF} from which temperature profile as well as stress-energy tensor can be obtained.

Let us consider the case where the quark starts at $(t,z) = 0$ 
and is dragged with constant velocity $\vec \beta=(\beta_x,\beta_y,\beta_z)$, meaning in particular that
the quark follows a trajectory whose $z$-component is 
$z=\beta_z t$. Along the trajectory of the quark, the fluid velocity 
is given by $v_z=z/t$, which is to say $v_z=\beta_z$, meaning 
that in the instantaneous fluid rest frame at all times the quark
is not moving in the $z$-direction.
The quark is moving
with the fluid in the $z$-direction.
In the laboratory frame, 
the fluid velocity gradients are given by
\beq
\begin{split}
%u^3 &= v_z u^0 = v_z \gamma_z \\
\partial_t u^3 &=  -\partial_z u^0 = -\gamma_v^3 \frac{\beta_z}{t} = - \gamma_v^2 \frac{\beta_z}{\tau_p}, \\
\partial_z u^3 &= - \frac{\partial_t u^0}{\beta_z^2} = \frac{\gamma_v^3}{t} = \frac{\gamma_v^2}{\tau_p},
\label{uderBF}
\end{split}
\eeq
where $\gamma_v \equiv (1-\beta_z^2)^{-1/2}$ is the relativistic gamma factor associated with the fluid velocity $\vec v$. We note that since the quark is in the local fluid rest frame at all times, 
the convective derivative of $u^3$ along the path of the quark vanishes: 
$D_t u^3=\partial_t u^3+ \beta^z \partial_z u^3 =0$.

By substituting \r{uBF} and \r{uderBF} into 
%either (\ref{forcesNotRF}) or
(\ref{f_final}), simplifying the resulting expression for $\vec f_{(1)}$, and combining
it with the zeroth order drag force (\ref{f0gen})
we find that
the drag force needed to pull the heavy quark at velocity $\vec\beta$ through
the Bjorken flow is given by
\beq
\begin{split}
\vec f_{\rm BF}(\tau_p) &=\vec f_{(0), \rm BF}(\tau_p) + \vec f_{(1),\rm BF}(t) =  \frac{\sqrt \lambda }{2\pi } \frac{\gamma}{\gamma_v\,b(\tau_p)^2} \( 1 + c_2\(-\frac{\gamma}{\gamma_v}\) \frac{b(\tau_p)}{\tau_p} \)
\left(
\begin{array}{c}
%    \gamma_z^2 \beta_\perp^2 \\
     \beta_{x} \\
     \beta_y \\
    \beta_z \gamma_v^2 \beta_\perp^2 
\end{array}
\right),
\end{split}
\label{forceBF}
\eeq
where $\beta_\perp^2 \equiv \beta_x^2 + \beta_y^2$ and where 
$c_2(-\gamma/\gamma_v)$ is defined in \r{cdef}, noting that for Bjorken Flow 
$s = -\gamma/\gamma_v$.  It is easiest to obtain (\ref{forceBF}) from (\ref{f_final}),
upon noting that since $u^3$ does not depend on $x$ or $y$
we have $u^\alpha \partial_\alpha u^3 = \gamma_v D_t u^3 =0$
and $w^\alpha \partial_\alpha u^3 = \gamma D_t u^3 =0$.  It can
also be shown that $\partial_z s=0$, meaning that the only 
nonvanishing term in (\ref{f_final}) is the term proportional to
$c_2(s)$.  At large $\tau_p$, $b(\tau_p)\sim \tau_p^{1/3}$ and
the first order term in (\ref{forceBF}) is smaller than the zeroth
order term by a factor $\sim \tau_p^{-2/3}$, which is the standard
power-counting for the derivative expansion in Bjorken flow.

When the quark is moving solely along the $z-$direction ($\beta_\perp = 0$), the drag force (\ref{forceBF}) 
vanishes identically at both zeroth and first order in gradients.   This is because in this case the quark is at rest in the 
instantaneous fluid rest frame at all times and, in the frame in which both the quark and the fluid
around it are at rest, there is no time derivative of the fluid velocity meaning that according
to (\ref{force_at_zero_speed}) there is no drag force.  So, in this case the existence
of fluid gradients does not modify the intuitive, zeroth order, result.
The result that we have obtained for the case
in which $\beta_\perp\neq 0$ and $\beta_z\neq 0$ looks less intuitive. 
However, note that it can be shown that if we boost the force (\ref{forceBF})
to the fluid rest frame, it has $f^z=0$ which means that in the fluid 
rest frame $\vec f \parallel \vec \beta$. If we choose $\beta_x\neq 0$ and $\beta_y=0$,
then in the fluid rest frame we find
\beq
f^x_{BF,RF} = \frac{\sqrt{\lambda}}{2\pi} \frac{1}{b(\tau_p)^2} \gamma \beta_x \(1 + \frac{b(\tau_p)}{\tau_p} c_2\(-\gamma\) \)\,.
\eeq
So, when $\beta_\perp\neq 0$ we find that the fluid gradients do correct
the result for the drag force at first order.

The drag force on a heavy quark moving through
a fluid expanding in a Bjorken flow 
has been discussed previously in the literature.
The leading term, namely $\vec f_{(0)}$ to zeroth order in gradients, was obtained 
in Refs.~\cite{Giecold:2009wi, Stoffers:2011fx}.
The authors of Ref.~\cite{Abbasi:2012qz} attempted the calculation
of the correction to the force to first order in fluid gradients for a heavy quark
moving through the Bjorken flow along a path with $z=0$ but, as we noted
previously, this calculation was based upon the assumption that the effects 
of fluid gradients could be attributed to their effects on the position of the
world-sheet horizon in the dual gravitational description, and we have 
now seen that the position of the world-sheet horizon is unaffected
by fluid gradients, at least to first order.

There are not many solutions to relativistic viscous hydrodynamics that are known
analytically.  Recently, Gubser has discovered two new such solutions, each
in a different sense a deformation of Bjorken flow.  
In the solution of Ref.~\cite{Gubser:2010ze}, the fluid expands in both
the transverse and longitudinal directions, with the longitudinal
expansion boost invariant as in Bjorken flow.  The other solution, in Ref.~\cite{Gubser:2012gy},
describes a longitudinal expansion that is not boost invariant but that can be obtained
via suitable deformation of Bjorken flow.  It would be interesting to apply our results
to obtain expressions for the drag force on a heavy quark moving through a fluid
expanding according to these hydrodynamic solutions. We leave this to future work.

\subsection{Colliding sheets of energy}
\label{sec:collsh}

We now return to the example that prompted our study~\cite{Chesler:2013cqa}, namely the drag force
needed to pull a heavy quark through the matter produced in the collision
of two planar sheets of energy in strongly coupled SYM theory, introduced 
in Ref.~\cite{Chesler:2010bi} and
analyzed there and in Refs.~\cite{Casalderrey-Solana:2013aba,Chesler:2013lia}.
The incident sheets of energy move at the speed of light in the $z$ and $-z$ directions and
collide at $z=0$ at time $t=0$.  They each have a Gaussian profile in the $z$ direction and
are translationally invariant in the two directions $\vec x_\perp=x,y$ orthogonal to $z$.
Because this setup is translationally invariant in $\vec x_\perp$ throughout the collision,
the motion of the fluid produced in the collision is entirely along the $z$ direction at all times.
The energy density per unit transverse area of the incident sheets is $\mu^3N_c^2/(2\pi^2)$
with $\mu$ an arbitrary scale with respect to which all dimensionful quantities
in the conformal theory can be measured.  As in Ref.~\cite{Chesler:2013cqa},
we shall choose the width $w$ of the Gaussian
energy density profile of each sheet to be $w=1/(2\mu)$.  Although there is no
single right way to compare the widths of these translationally invariant sheets
of energy with Gaussian profiles to the widths of a nucleus that has
been Lorentz-contracted by a factor of 107 (RHIC) or 1470 (LHC), reasonable
estimates suggest that our choice of $w\mu$ corresponds to sheets with
a thickness somewhere between the thickness of the incident nuclei at  RHIC and LHC~\cite{Chesler:2010bi}.
%
%
%The collision of two shockwaves with finite thickness in the asymptotically AdS$_5$ was originally formulated in \cite{Chesler:2010bi} and further investigated in \cite{Casalderrey-Solana:2013aba}. 
%
The matter produced in these collisions is initially far from equilibrium but it then rapidly
hydrodynamizes: after a time $t_{\rm hydro}$ its subsequent expansion and cooling
is well described by viscous hydrodynamics, with $t_{\rm hydro}/ b(t_{\rm hydro})$ at most
2-3~\cite{Chesler:2010bi}.  
%
%The key aspect in these collisions is that at time corresponding to 2-3 sheet thickness after the collision the fluid reaches equilibrium and first order hydrodynamics become applicable in local patches. 
%

In Ref.~\cite{Chesler:2013cqa} we and a coauthor inserted a heavy quark
moving with velocity $\vec \beta$ between the colliding sheets before
the collision, choosing a trajectory such that the heavy quark is
at $z=0$ at $t=0$, meaning that it finds itself in the center of
the collision, and calculated the drag force needed to keep
the velocity of the heavy quark constant throughout the collision.
Our focus throughout much of Ref.~\cite{Chesler:2013cqa} was 
the drag force at the earliest moments of the collision when the
matter was far from equilibrium.  We also calculated the drag force
during the later epoch when the fluid has hydrodynamized and is expanding
according to first order viscous hydrodynamics.    We compared our results
throughout to expectations for what the drag force would have been in 
a spatially homogeneous equilibrium fluid with the same instantaneous energy density,
transverse pressure or longitudinal pressure.  The first of these corresponds
to the zeroth order drag force (\ref{f0gen}).    To see this, note that what we 
did in Ref.~\cite{Chesler:2013cqa} was to first boost to the instantaneous fluid
rest frame, then compute the energy density $\varepsilon_{RF}$ in that frame, and from that define
a temperature $T_e$ as if the fluid were spatially homogeneous and
in equilibrium, which is to say via $\varepsilon_{RF}=3\pi^2N_c^2 T_e^4/8$,
and then use this $T_e$ in the expression for the drag force on a heavy quark moving
through an equilibrium fluid with no gradients.  From (\ref{Tmunu}) and (\ref{sigmamunu}) we see
that in the instantaneous fluid rest frame $\sigma^{00}$ vanishes, meaning that
in this frame the fluid gradients do not contribute to $T^{00}=\varepsilon_{RF}$.
Thus, the $T_e$ we defined in Ref.~\cite{Chesler:2013cqa} is related to $b$
precisely by $b=1/(\pi T_e)$.  So, the dashed curves in Ref.~\cite{Chesler:2013cqa}
that were drawn using $T_e$ correspond precisely to expectations for the
drag force upon working to zeroth order in fluid gradients, namely (\ref{f0gen}).
The results of Ref.~\cite{Chesler:2013cqa} can be summarized as follows.
First, (\ref{f0gen}) has roughly the right magnitude even just
after the collision when the matter is far-from-equilibrium, although the time dependence 
of the actual
force lags behind that obtained via  (\ref{f0gen}) by a time delay that grows linearly
with increasing $\gamma$.  And, second, 
it was noted in Ref.~\cite{Chesler:2013cqa} that even after the fluid has hydrodynamized
the actual drag force calculated there does not agree with (\ref{f0gen}), an effect
that was attributed to the effects of gradients in the fluid.  Here we shall confirm
this attribution.

We shall compare the drag force calculated in the full calculation of Ref.~\cite{Chesler:2013cqa}
to the zeroth order expectation (\ref{f0gen}), which neglects
the effects of fluid gradients, and to that plus the contribution due to fluid
gradients to first order which we now have at our disposal in the form (\ref{forcesNotRF}) or the form (\ref{f_final}). 
We shall do the comparison for two cases in which $\beta_z=0$ and $\beta_x\neq 0$, meaning
that the quark is moving perpendicular to the fluid motion,  two cases in which
$\beta_x=0$ and $\beta_z\neq 0$, with the quark moving in the same direction as the fluid,
and one case in which both $\beta_x$ and $\beta_z$ are nonzero.

\begin{figure*}[t]
\includegraphics[scale=0.40]{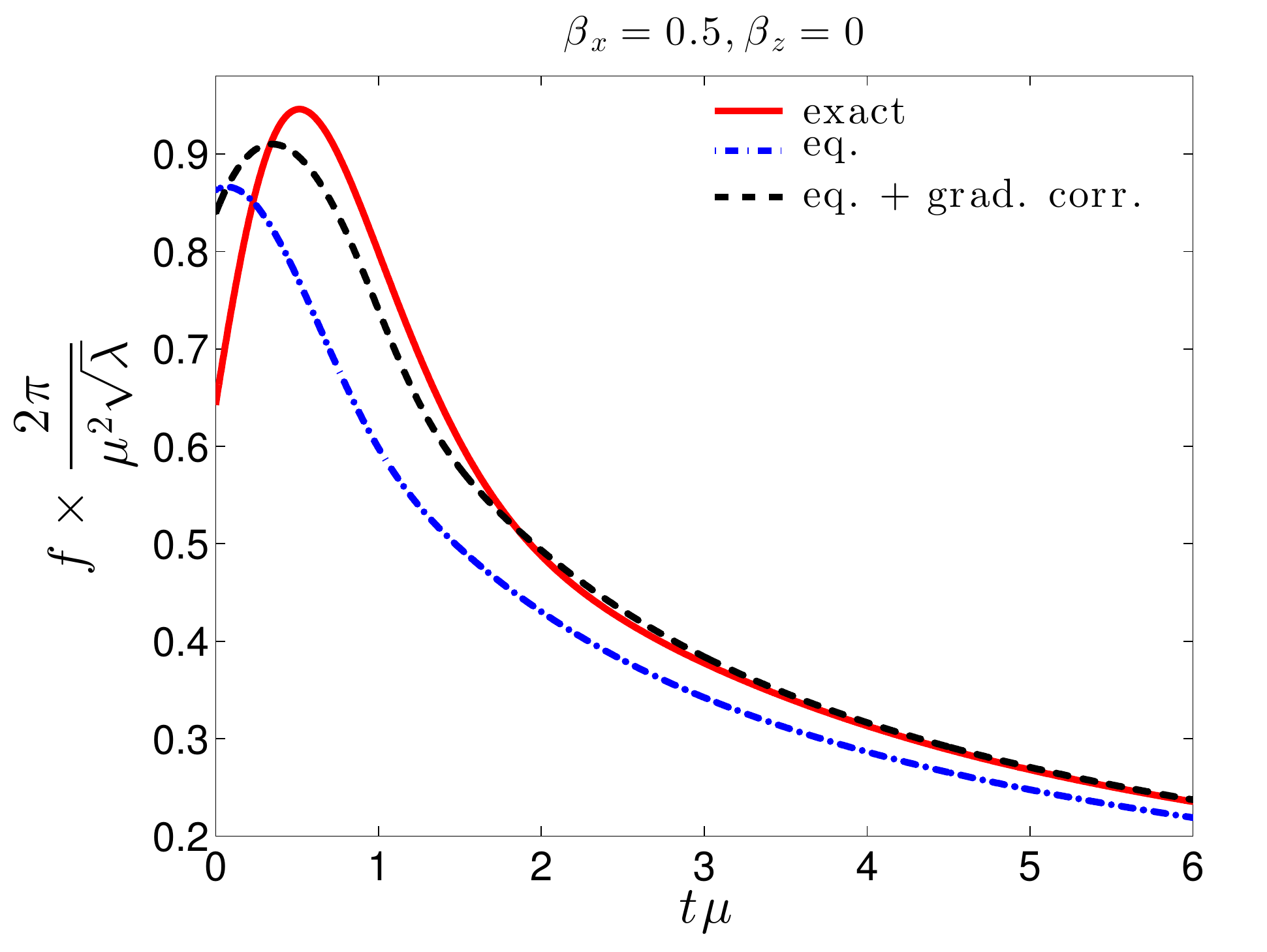}
\includegraphics[scale=0.40]{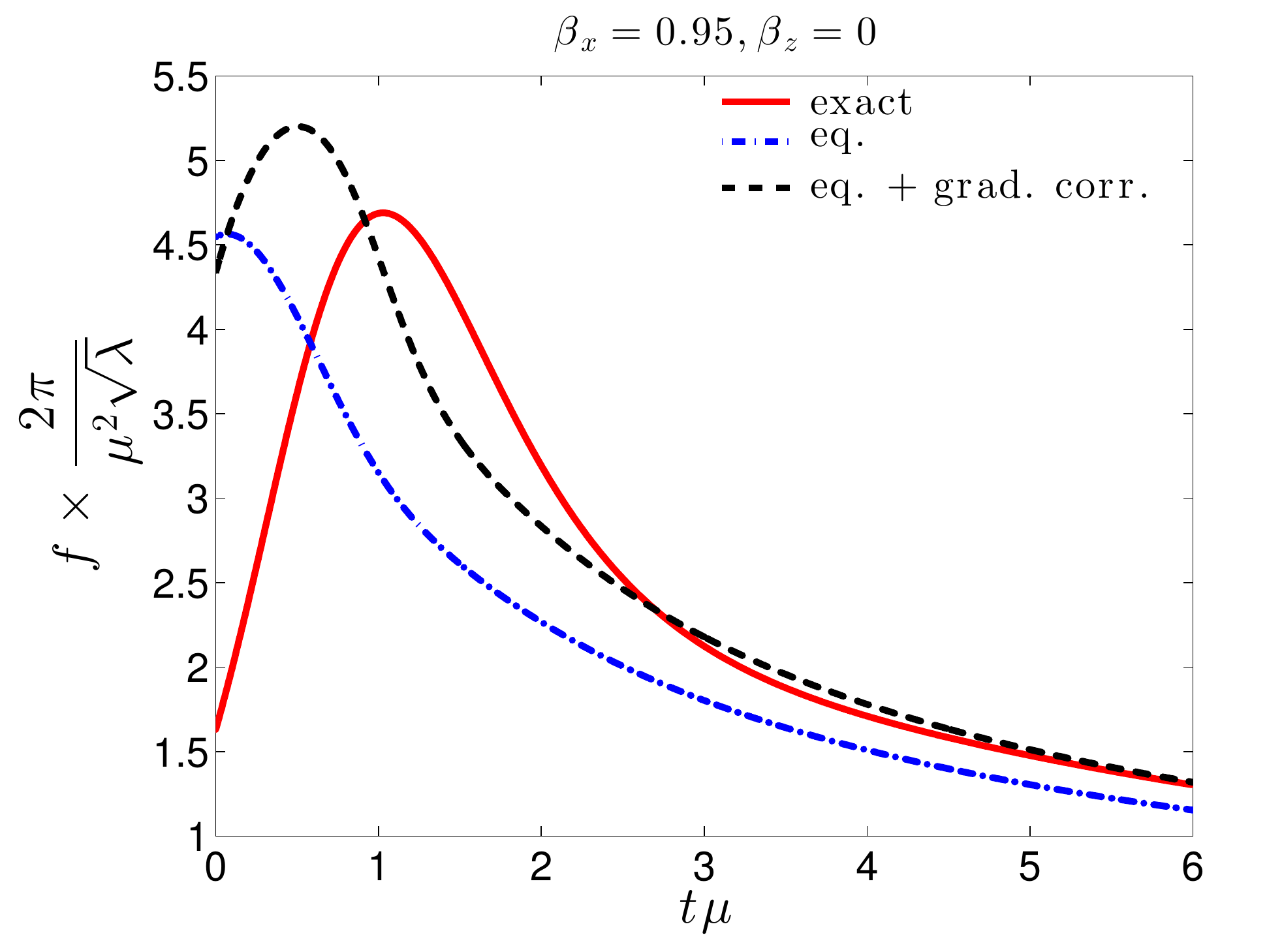}
\caption{Actual drag force (red curves) from Ref.~\cite{Chesler:2013cqa} 
on a heavy quark being dragged with $\beta_z=0$
and $\beta_x=0.5$ (left panel)  or $\beta_x=0.95$ (right panel)
through the debris produced in the collision of two sheets of energy.
We compare the actual drag force to the zeroth order calculation (blue dot-dashed curve)
which neglects the effects of fluid gradients and our 
calculation in which the effects of fluid gradients are
included up to first order (black dashed curves).
%compared to our zeroth order calculation (blue dashed curve)
%comparison with the equilibrium expectation drag force (dot-dashed blue line) and with the gradient corrections included (black dashed line) when the quark moves along the zero rapidity trajectory ($z=0$). For drag speed $\beta_x = 0.5$ (left panel) as well as for $\beta_x = 0.95$ (right panel), the gradient corrected eq. drag force agrees with the exact result with small differences caused by the second order fluid gradients.
At late times, when the fluid has hydrodynamized, the gradient corrections
included in the black dashed curves yield a much better description of the
full result.
}
\label{zeroRap}
\end{figure*}

In Fig.~\ref{zeroRap} we plot the drag force on a quark moving in the $x$-direction, perpendicular
to the ``beam'' direction and therefore perpendicular to the direction of motion of the fluid,
with $\beta_x=0.5$ and $\beta_x=0.95$.  The red curves show the drag force
obtained from the full gravitational calculation of Ref.~\cite{Chesler:2013cqa}, without
any expansion in gradients.  The blue dot-dashed curves, which were also obtained
in Ref.~\cite{Chesler:2013cqa}, so what the drag force would be at each instant
in time in a static homogeneous fluid in thermal equilibrium with the same energy density as
the actual fluid has at that instant in time at the location of the quark.  An equivalent
description of these curves, which are obtained from our expression
(\ref{f0gen}) that is zeroth order in fluid gradients, is that they show what the drag force would be if we neglect all effects
of the spatial gradients and variation in time of the fluid at the location of the quark.
The black dashed curves show how the drag force changes when we
start with the blue dot-dashed curves and add the results of our calculation (\ref{f_final}) of
the first-order effects of fluid gradients on the drag force.  
Using the operational definition of the hydrodynamization time $t_{\rm hydro}$ introduced
in Ref.~\cite{Chesler:2010bi}, namely taking it to be the time after which the transverse
and longitudinal pressure agree with those obtained via the hydrodynamic constitutive
relations from the energy density and the fluid velocity, in 
Fig.~\ref{zeroRap} hydrodynamization time $t_{\rm hydro}\mu = 2.8$.
We see that after $t_{\rm hydro}$  the black dashed curves
are much closer to the full results shown by the red curves than the blue dot-dashed curves are, 
meaning that adding effects of fluid gradients to first order has resolved
most of the discrepancy between the full results and the zeroth order blue dot-dashed curves.
This confirms that this discrepancy was due to the effects of the fluid gradients.  It is
reasonable to guess that if one were to push our calculation to second order in gradients,
the agreement would get even better. We leave this to future work.

We have also checked that the criteria (\ref{limit_gamma}) are well satisfied, by more than 
a factor of two in fact, at all times after $t_{\rm hydro}$ even for $\beta_x=0.95$,
namely for $\gamma=3.2$.  Throughout, we will only
show results for cases in which these criteria are satisfied by a large margin.

\begin{figure*}[t]
\includegraphics[scale=0.41]{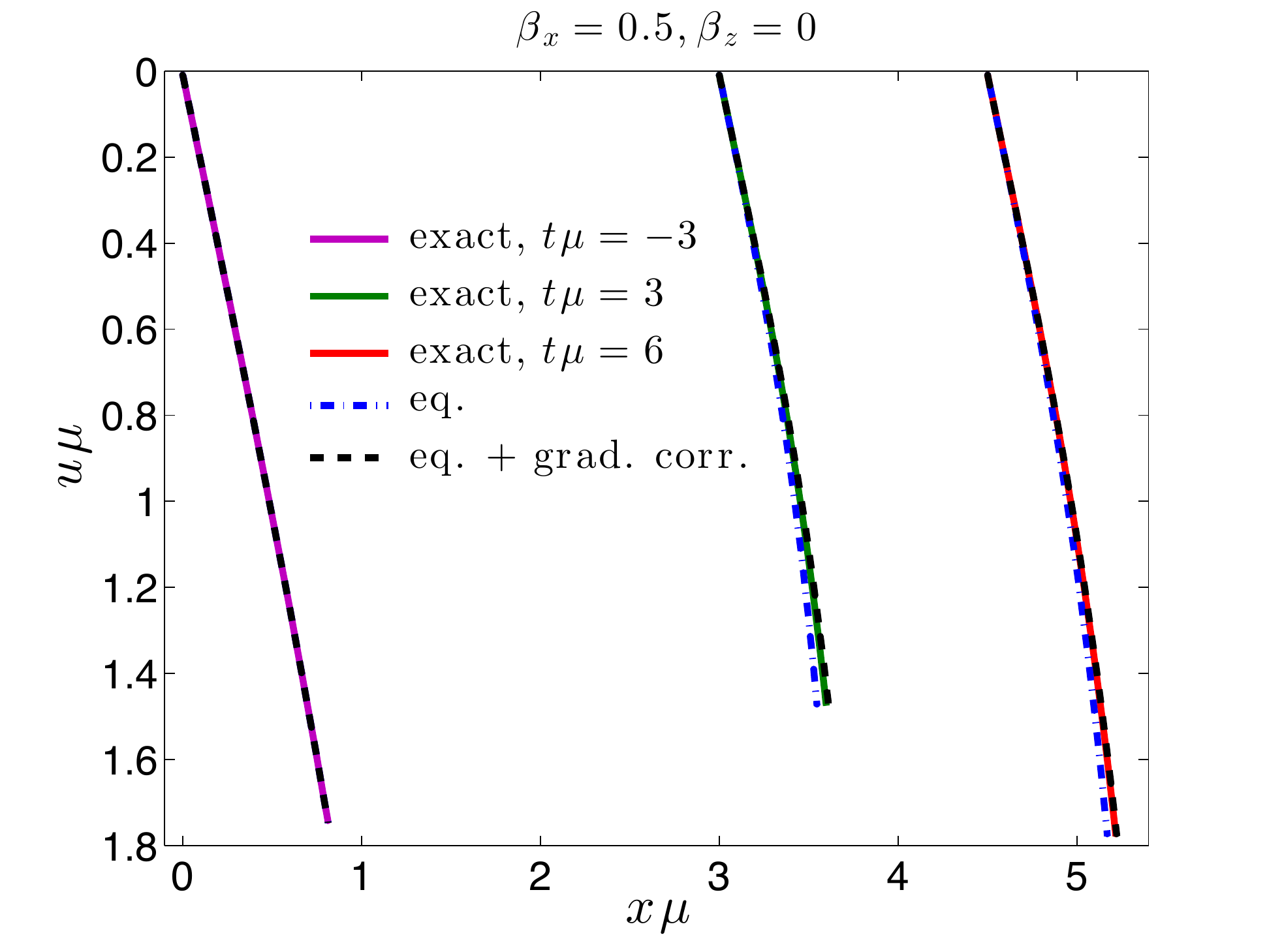}
\includegraphics[scale=0.41]{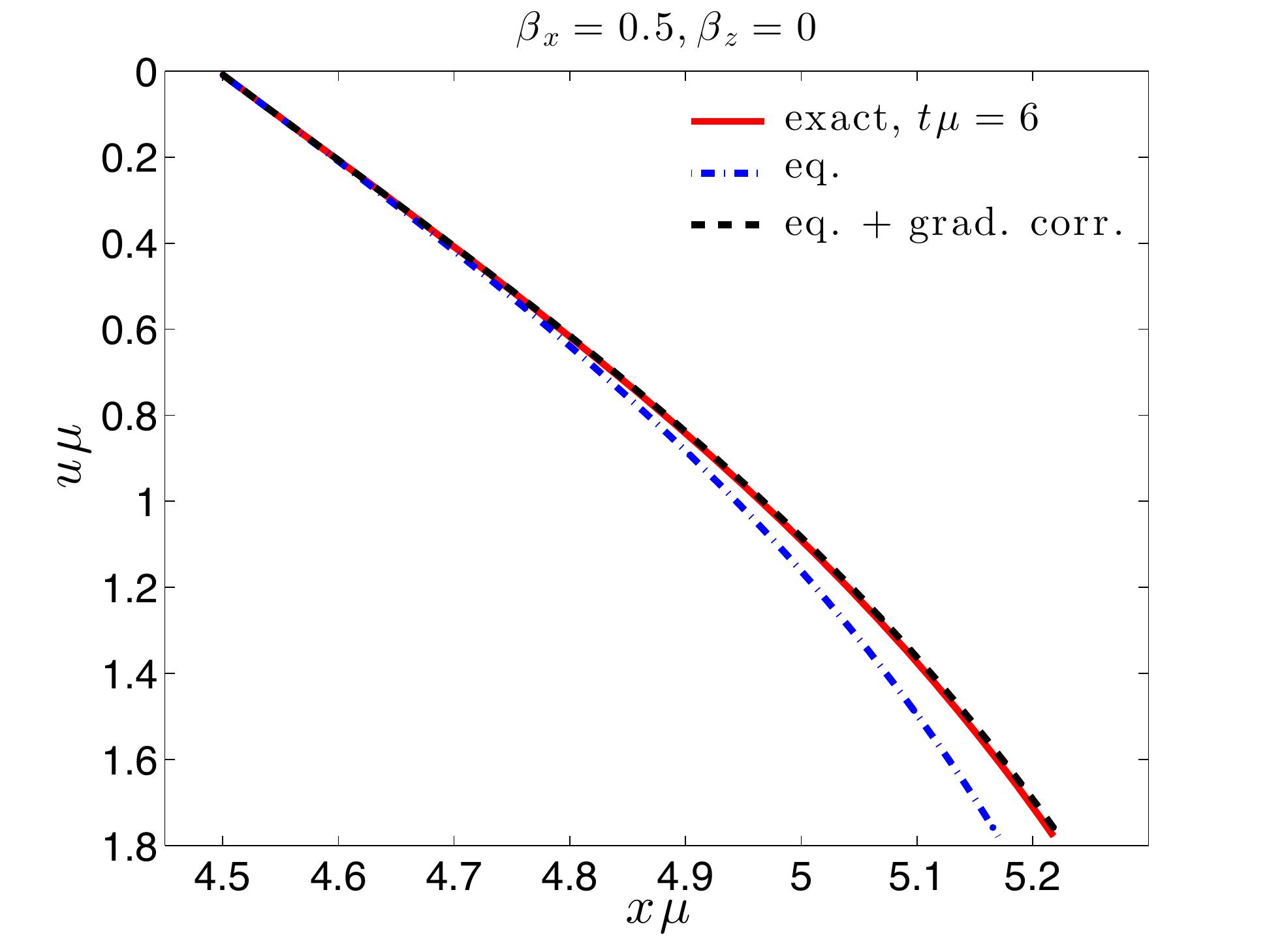}
\caption{Comparison of the profile of the string trailing ``down'' into the bulk from the 
heavy quark moving with $\vec\beta=(0.5,0,0)$.  The vertical axes show the
radial, or holographic, coordinate $u=1/r$,
meaning that the AdS boundary at $r=\infty$ is at $u=0$, at the top of the figures.
The horizontal axes show $x$; the quark and hence its string is 
moving to the right.  
The curves show the shape of the string
at a fixed Eddington-Finkelstein time $t$.
The left panel shows the string at three times, $t\mu=-3, 3$ and 6.  The right panel
zooms in at $t\mu=6$.  In all cases, the solid curve shows the string profile
obtained from the full gravitational calculation in Ref.~\cite{Chesler:2013cqa},
the  blue dot-dashed curve shows the string profile (\ref{x0gen}) as it would be
at that instant in time $t$ if gradients in the fluid were neglected,
and the black dashed curve shows the string profile (\ref{x_profile2})
including the effects of fluid gradients to first order.
}
\label{betax05shape}
\end{figure*}

To get further intuition, in Fig.~\ref{betax05shape} 
we investigate the shapes of the string hanging ``down'' into the
gravitational spacetime from the heavy quark
in the calculation of the drag force shown in the left panel of Fig.~\ref{zeroRap}.
%and \ref{betaz02shape}, 
Each string profile is plotted at fixed Eddington-Finkelstein 
time $t\mu$ as a function of the inverse radial coordinate $u = 1/r$.
The solid curves
are the exact string profiles at three times,
obtained numerically in the gravitational calculation of  Ref.~\cite{Chesler:2013cqa}.\footnote{
The drag force is independent of one's choice of coordinates for the 4+1-dimensional 
gravitational metric, but when we plot the shape of the string $u(x)$ at one value
of the time coordinate $t$ this shape does of course depend on one's definition of the coordinates $u$ and $t$.
The calculation in Ref.~\cite{Chesler:2013cqa} was done using a metric in which $G_{tr} = 1$
and $G_{Mr}$ vanishes for $M\neq t$. In our calculation of \r{gs} we have instead 
used the metric given in Eqs.~(\ref{GMN}, \ref{GMN0gen}, \ref{GMN1gen}) in which
$G_{rr} = 0$ and  $G_{\mu r} = -u_\mu$.  In order to make the comparison in 
Fig.~\ref{betax05shape}, we have transformed the exact results for the string
profile obtained in Ref.~\cite{Chesler:2013cqa} from the metric used there
to the metric we are using here.
This coordinate transformation can be determined
order-by-order in the fluid gradients, as described in Ref.~\cite{Chesler:2013lia}.
}
The blue dot-dashed curves are zeroth order
in fluid gradients: they show the shape (\ref{x0gen}) that the
string would have in a static, spatially homogeneous, equilibrium
fluid with the same energy density as that at the location of the heavy quark.
The black dashed curves are obtained by integrating $\partial_\sigma \vec g$
in the fluid rest frame, {\it i.e.}, Eqs.~(\ref{gs}).  In a case like those we shall
turn to below, where the lab frame is not the same as the
fluid rest frame, we would then boost the string profile from the
fluid rest frame back to the lab frame.
We see from Fig.~\ref{betax05shape} that including the effects of
fluid gradients on the string profile to first order yields a much better
description of the actual string profile, just as for the
drag force itself.

\begin{figure*}[t]
\includegraphics[scale=0.40]{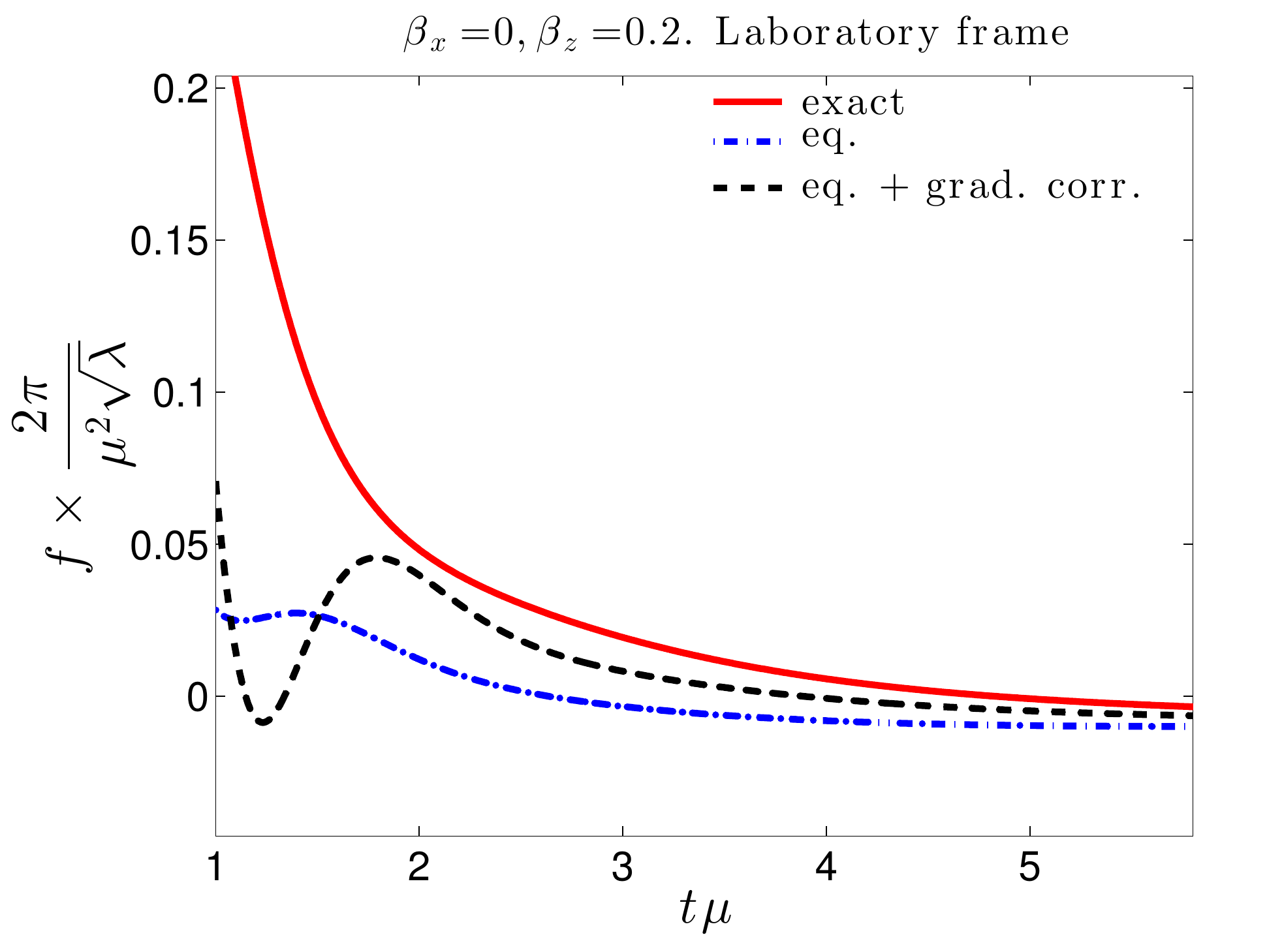}
\includegraphics[scale=0.40]{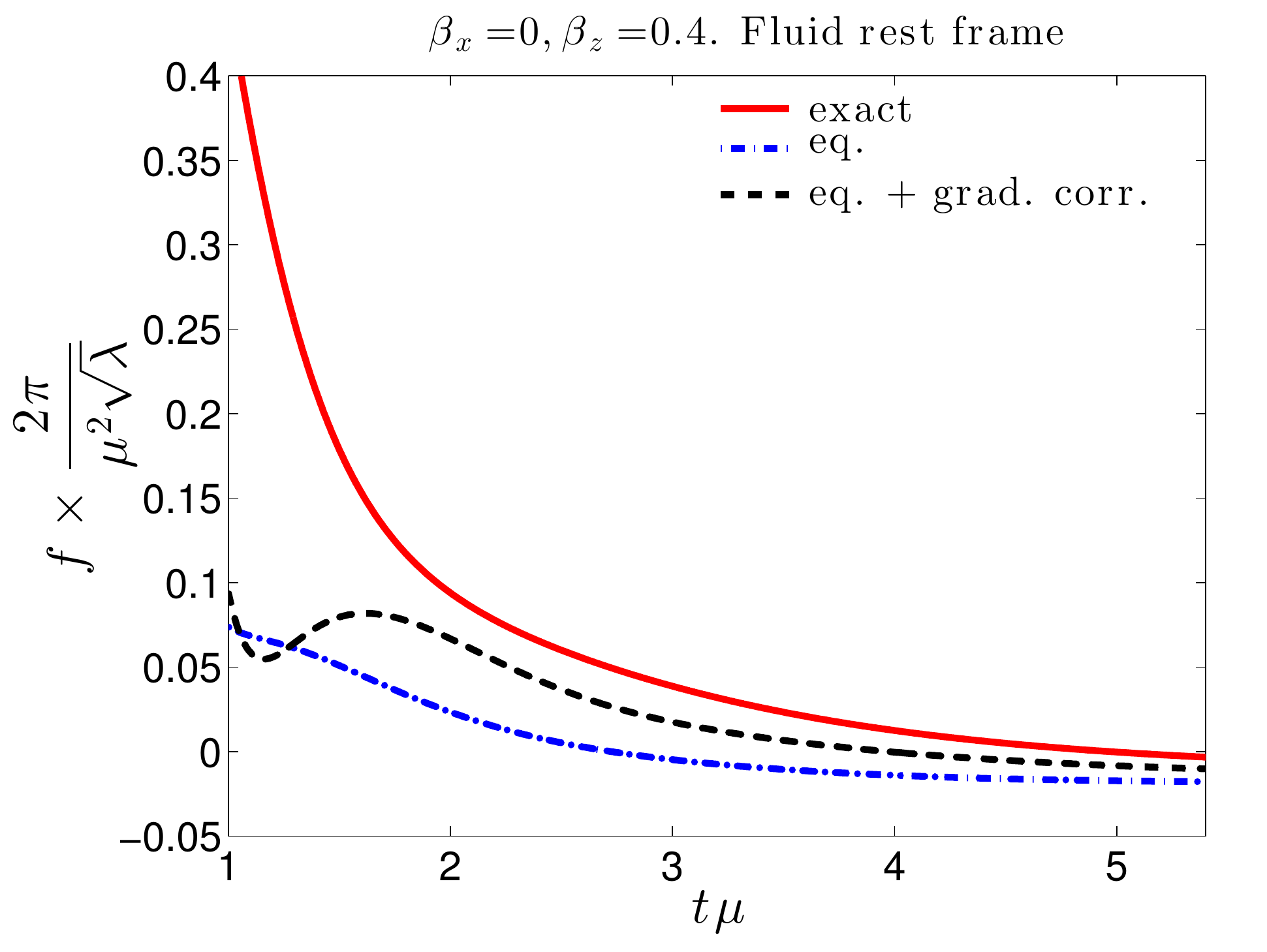}
\caption{As in Fig.~\ref{zeroRap}, except that here the quark has zero velocity in the
direction perpendicular to the motion of the fluid and is moving only in the $z$-direction.
In the left panel, $\beta_z=0.2$ and in the right panel $\beta_z=0.4$.
We have shown the left panel in the laboratory frame while in the right panel
at each time $t$ we have boosted to the frame in which the fluid is at rest at
the location of the quark.
As in Fig.~\ref{zeroRap}, we show the exact results for the drag
force obtained in Ref.~\cite{Chesler:2013cqa} as well as the
zeroth-order approximation (i.e. the drag force in a static homogeneous fluid
with the same instantaneous energy density) and the result
that we have obtained upon including the effects of fluid gradients
to first order.
}
\label{nonzeroRap}
\end{figure*}

In Fig.~\ref{nonzeroRap} we investigate two cases when the 
quark is moving with nonzero rapidity, $\beta_z\neq 0$.  Here we
choose to set $\beta_x=0$; below we will consider a case when
both $\beta_x$ and $\beta_z$ are nonvanishing.  
In Fig.~\ref{nonzeroRap} we have chosen 
$\beta_z = 0.2$ and $\beta_z = 0.4$. In both cases, and as in Fig.~\ref{zeroRap}, 
including the first order effects of fluid gradients on the drag force improves
the agreement with the exact calculation of the drag force from Ref.~\cite{Chesler:2013cqa}.
In Fig.~\ref{nonzeroRap} the agreement between the black dashed curves
and the solid red curves is worse than in Fig.~\ref{zeroRap}, in fractional terms,
but the more striking difference between the two Figures is that the overall magnitude
of the forces plotted in Fig.~\ref{nonzeroRap}
is more than an order of magnitude smaller than the forces in Fig.~\ref{zeroRap}.
This can be understood by recalling our results for Bjorken flow, from Section \ref{sec:bj}.
If the longitudinal expansion of the fluid produced in the collision that we are analyzing
here were boost invariant, our results from Section \ref{sec:bj} tell us that
when we choose $\beta_z\neq 0$ and $\beta_x=0$ we would find
no drag force at all, at zeroth and first order in fluid gradients.  The fact that
we see a nonzero drag force in Fig.~\ref{nonzeroRap} reflects the fact
that the expansion of the fluid produced in the collision is not boost invariant.
Since at late times the expansion is close to boost invariant~\cite{Chesler:2013lia},
all the forces in Fig.~\ref{nonzeroRap} are small in magnitude.  Upon realizing this,
we also note that the absolute difference between the black dashed and
solid curves in Fig.~\ref{nonzeroRap} is in fact quite similar to their absolute difference 
in Fig.~\ref{zeroRap},
meaning that the larger fractional deviation in Fig.~\ref{nonzeroRap} is simply an
artifact of the smallness of the magnitude of the drag force which is a consequence
of the expansion being almost boost invariant.

\begin{figure*}[t]
\begin{center}
\includegraphics[scale=0.41]{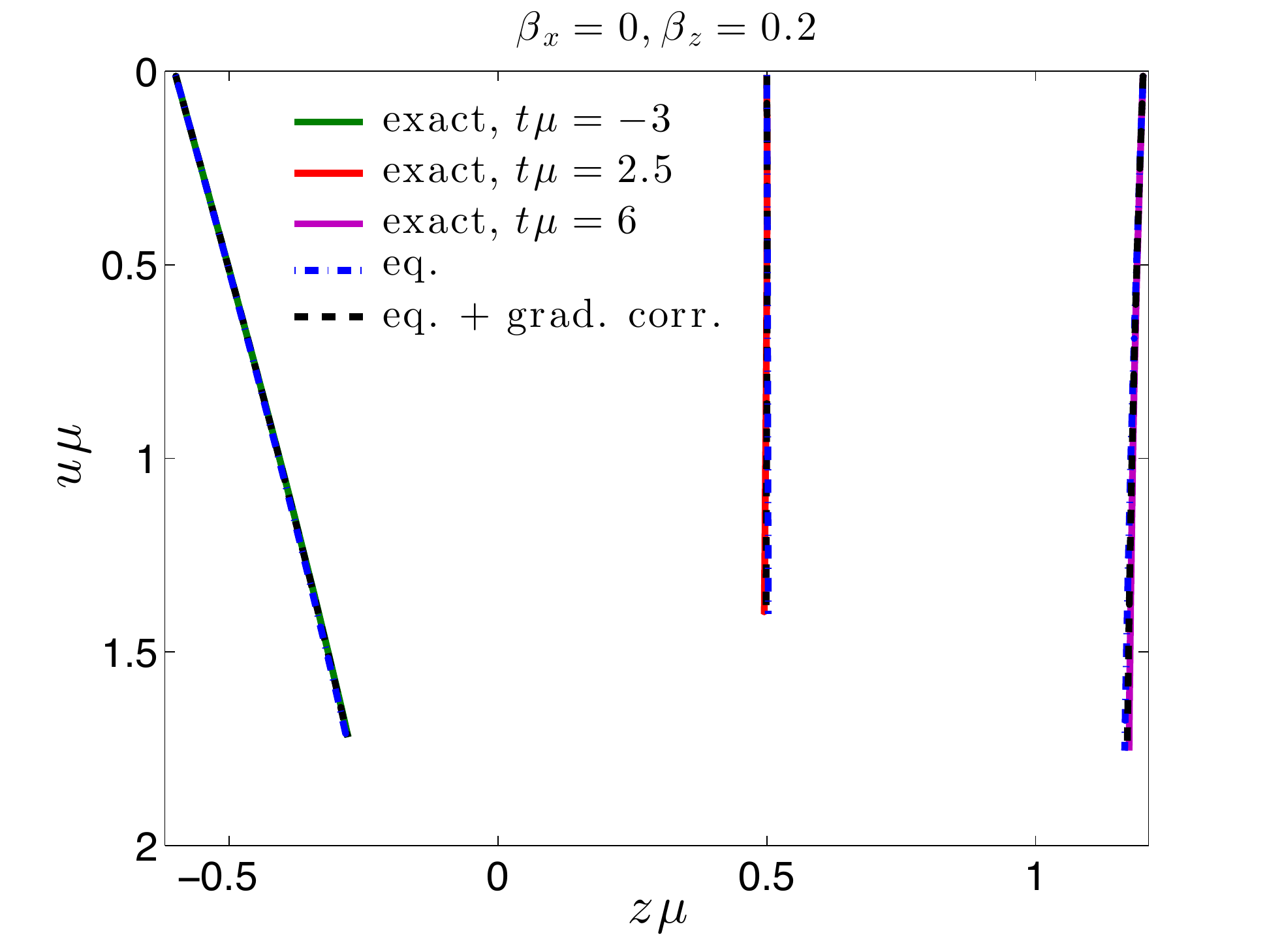}
\end{center}
\vspace{-0.15in}
\includegraphics[scale=0.41]{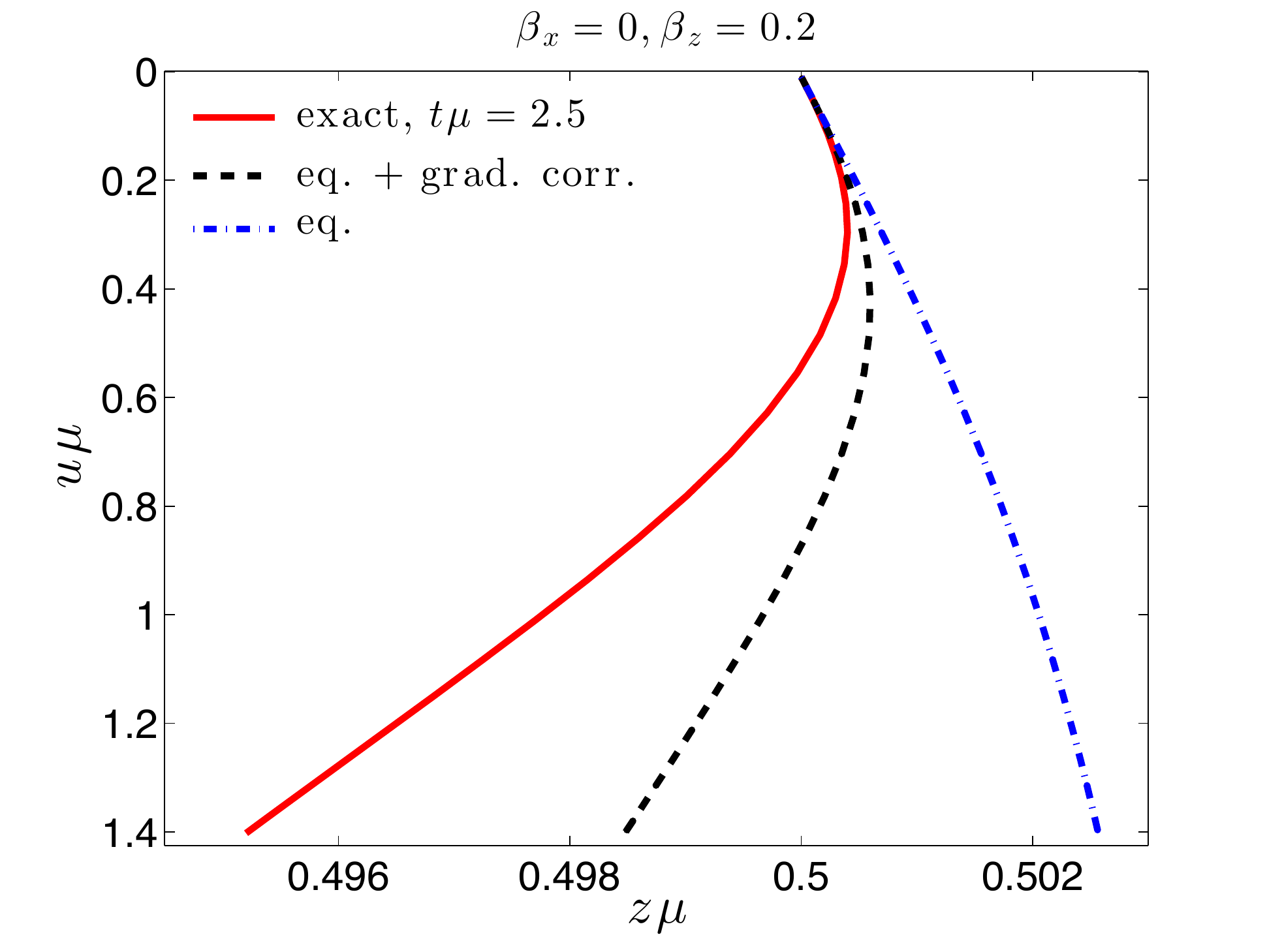}
\includegraphics[scale=0.41]{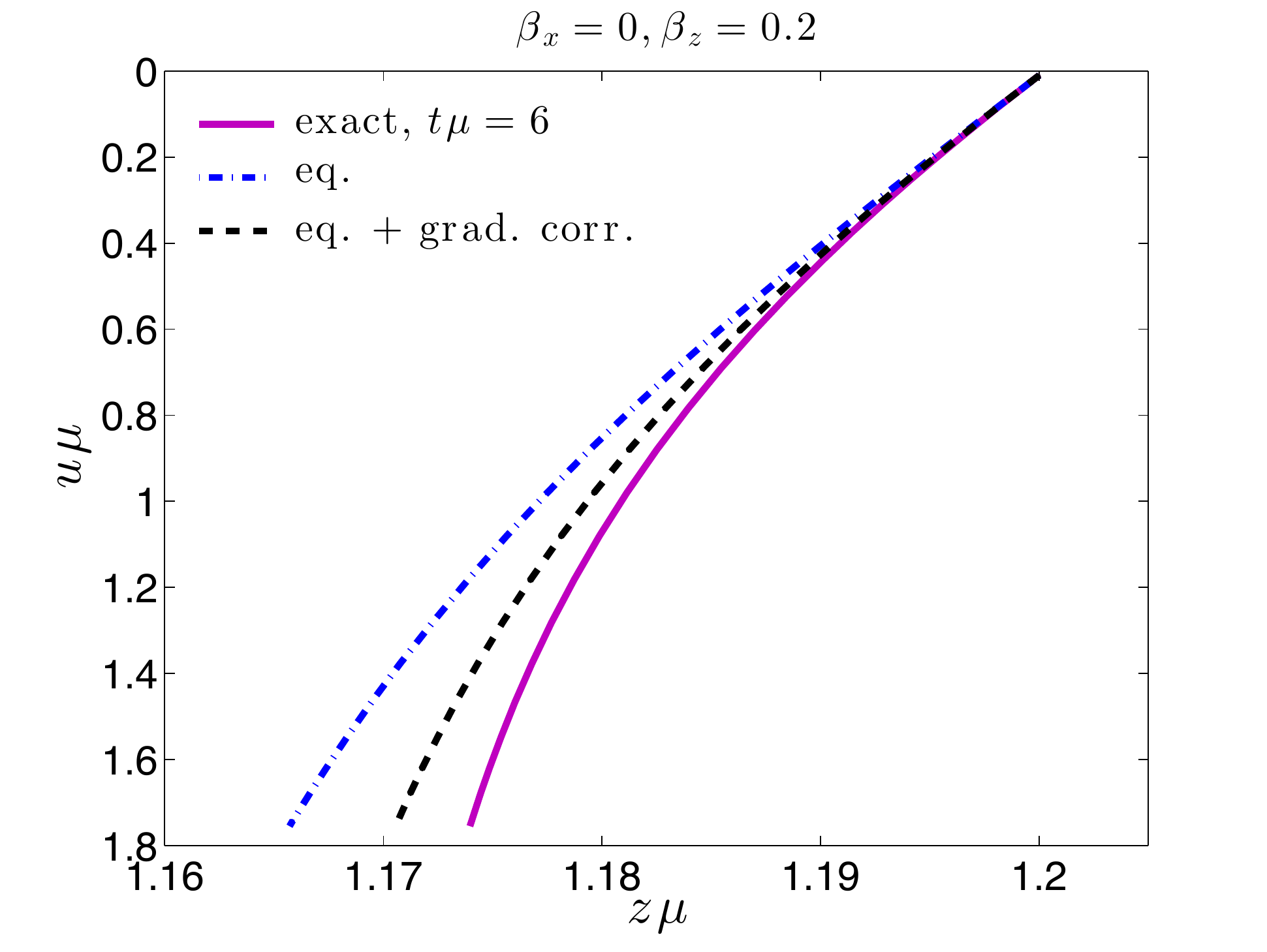}
\caption{Comparison of the profile of the string trailing ``down'' into the bulk from the 
heavy quark moving with $\vec\beta=(0,0,0.2)$.  The axes
are as in Fig.~\ref{betax05shape}.
The solid curves show the shape of the string
obtained from the full gravitational
calculation of Ref.~\cite{Chesler:2013cqa}
at three fixed Eddington-Finkelstein times $t$, namely
$t\mu=-3, 2.5$ and 6.
As in Fig.~\ref{betax05shape}, the blue dot-dashed curves show the string profile as if there were no gradients
in the fluid and the black dashed curves show the results of this paper, with the 
effects of  fluid gradients taken into account to first order.
The lower panels zoom in on the string profiles at $t\mu=2.5$ and 6.  We have chosen
$t\mu=2.5$ as one of the times at which we illustrate the string profile because it is
just before the time $t\mu=2.63$ at which the blue dot-dashed curve in the left panel
of Fig.~\ref{nonzeroRap} crosses zero, which is to say it is just before the time at
which the relative velocity of the quark and the fluid changes sign, meaning that
the
zeroth-order estimate of the drag force changes sign.
}
\label{betaz02shape}
\end{figure*}

In Fig.~\ref{betaz02shape} we investigate the shapes of the string
attached to the heavy quark moving with $\beta_z=0.2$ whose drag force is shown in the 
left panel of Fig.~\ref{nonzeroRap} at the three Eddington-Finkelstein
times
$t\mu = -3, 2.5$, and $6$. 
As in Fig.~\ref{betax05shape}, we see that including the effects of
fluid gradients on the string profile to first order improves the
description of the exact string profile obtained in Ref.~\cite{Chesler:2013cqa}.
Just as when we compared Figs.~\ref{zeroRap} and \ref{nonzeroRap}, when
we compare the zoomed-in panels of Fig.~\ref{betaz02shape} to the zoomed-in
panel of Fig.~\ref{betax05shape} we see that the absolute differences between
the analytic results to first order in fluid gradients (black dashed curves)  
and the full results obtained numerically (solid curves) are comparable, although the
fractional deviations look greater in Fig.~\ref{betaz02shape}.

\begin{figure*}[t]
\includegraphics[scale=0.4]{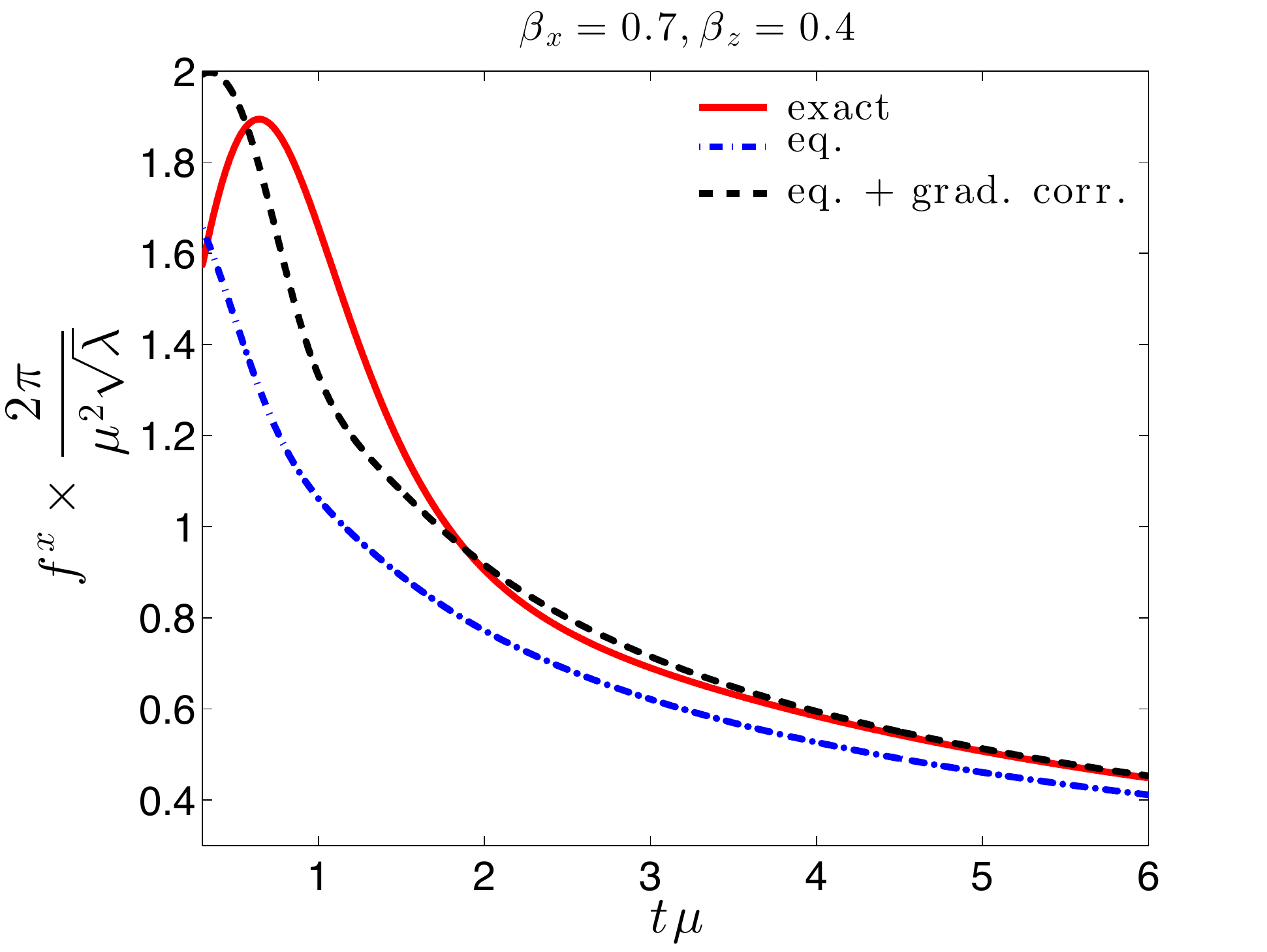}
\includegraphics[scale=0.4]{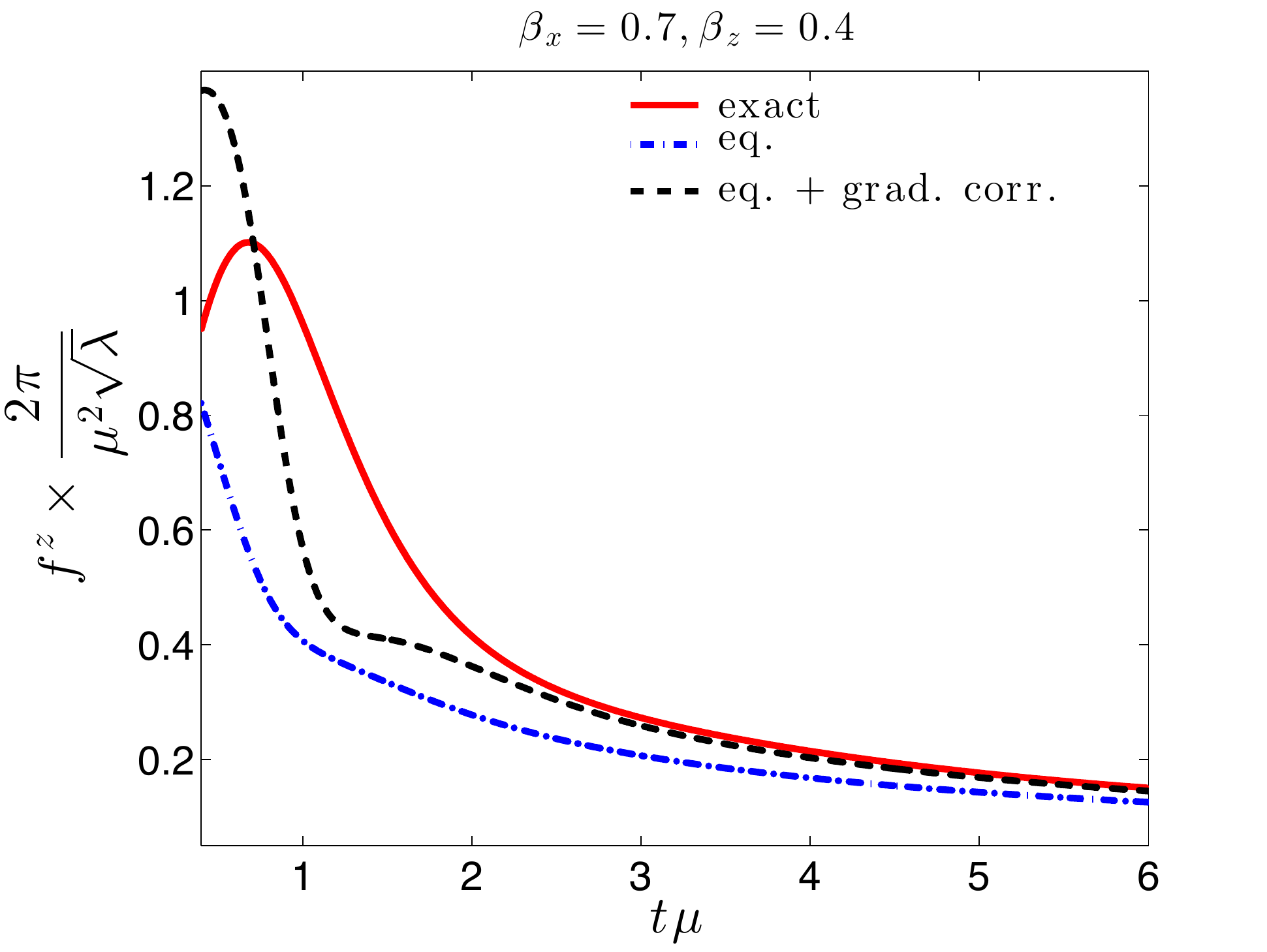}
\includegraphics[scale=0.39]{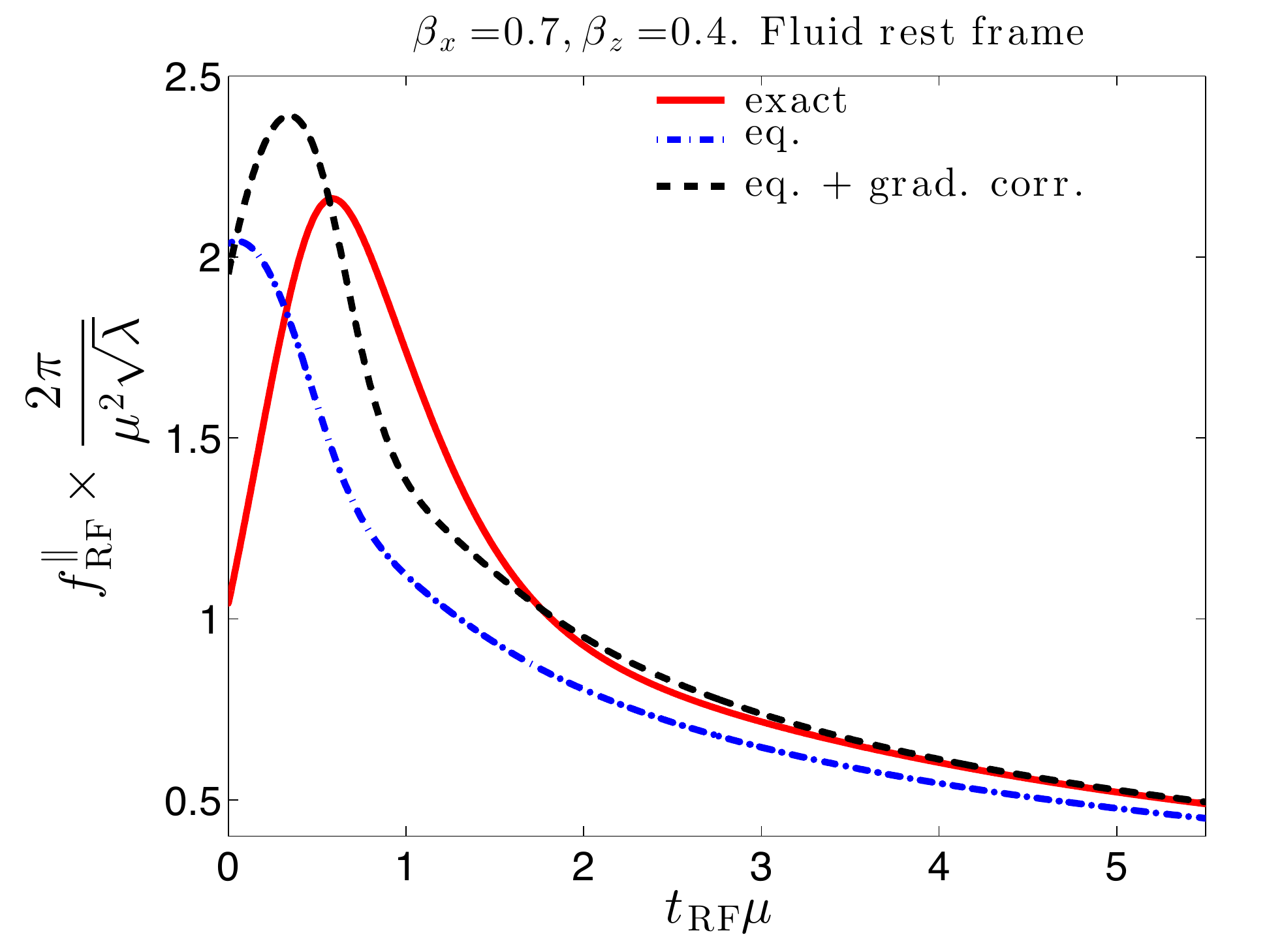}
\includegraphics[scale=0.39]{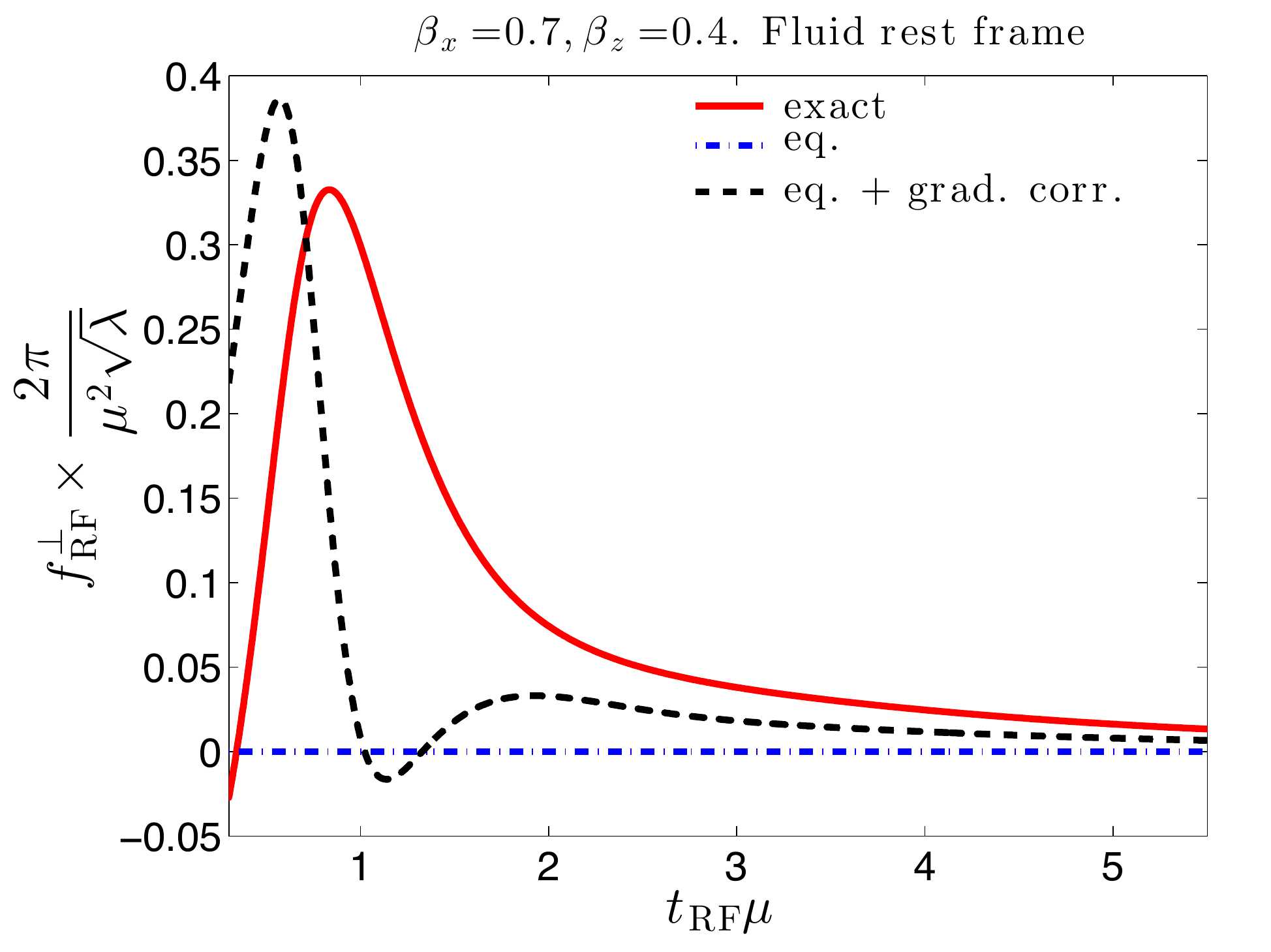}
\caption{As in Fig.~\ref{zeroRap}, but the case when quark is moving with velocity
$\vec\beta=(0.7,0,0.4)$, meaning 
 $\beta_x = 0.7$ perpendicular to the collision direction and $\beta_z = 0.4$ along the collision axis. 
 The upper two panels show the 
 $x$- and $z$-components (top-left and top-right panels, respectively) of the drag force as seen in the
``laboratory frame'', which is to say the center-of-mass frame for the collision.
In the lower two panels, we boost to the frame in which the fluid at the location
of the heavy quark is at rest. And, instead of showing the $x$- and $z$-components
of the drag force in this frame, we show the components of the force in the
directions parallel to (bottom-left panel) and perpendicular to (bottom-right panel)
the direction of motion of the quark in the local fluid rest frame.
In all the panels, the drag force with fluid gradient corrections included
to first order (black dashed curve) gives a better description of the 
full drag force (red curve) than does the zeroth-order drag force with fluid gradients neglected.
}
\label{betax07betaz04}
\end{figure*}

We have chosen $t\mu=2.5$ as one of the times at which
we illustrate the string profile in Fig.~\ref{betaz02shape} because it is close to the time $t\mu=2.63$
at which the velocity of the fluid at the location of the quark, $v_z$, goes from
below $0.2$ to above $0.2$, meaning that the relative velocity of the quark
and the fluid changes sign at that time.  At $t\mu=2.63$, the zeroth-order approximation
to the drag force therefore changes sign, as seen in the blue dot-dashed curve
in the left panel of Fig.~\ref{nonzeroRap}.  We see that this change is also reflected
in the string profile: at $t\mu=2.63$, the string would be hanging straight down
from the quark at the boundary; earlier, it angles to the right; later, it angles to the
left.  We see that at $t\mu=2.5$ the orientation of the string has already changed deeper
within the bulk and the change in orientation is about to reach the boundary.
Note that the orientation of the string at the boundary suffices to determine
the sign of the drag force only to zeroth order.  Once the effects of fluid
gradients are included, the drag force at time $t$ depends on how the string is moving
as well as on the orientation of the string~\cite{Chesler:2013cqa}.   
We see in the left panel of Fig.~\ref{nonzeroRap}
that the drag force including effects of fluid gradients to  first order 
(black dashed curve) and the full drag force (red curve) change sign only considerably
later than $t\mu=2.63$.  
%\color{red}{
Starting at $t\mu=2.63$, when the
relative direction of the fluid flow and the quark changes, 
we have a period of time when the drag force exerted
by the fluid on the quark points in the same direction
as the velocity of the quark, an effect
that was highlighted in Ref.~\cite{Chesler:2013cqa}.
We now see from the black dashed curve that this effect
can be accounted for qualitatively by
the effects of fluid gradients to first order.

The $t\mu=2.5$ panel of Fig.~\ref{betaz02shape} is also interesting
insofar as it shows an example where although the difference between the zeroth order
string profile and the full string profile is small in magnitude these two profiles
have qualitatively different shapes, and we see the first order effects of gradients 
doing the job of turning the blue dot-dashed curve into a black dashed curve that
looks much more like the red solid curve.

Finally, in Fig.~\ref{betax07betaz04} we show the results of our calculation of the
effects of fluid gradients to first order on the drag force needed to move a heavy 
quark along a trajectory with both $\beta_x\neq 0$ and $\beta_z\neq 0$.
The message from the upper two panels is much the same as what we have
already learned from Fig.~\ref{zeroRap}.  In the lower two panels, at each time we boost
to a frame in which the fluid at the location of the heavy quark is instantaneously at rest.  In this
frame, the heavy quark is of course still moving, with a substantial velocity in the $x$-direction
and some velocity in the $z$-direction.   We have chosen to plot the components of the drag force
in this frame in the directions parallel to and perpendicular to the direction of motion
of the quark in this frame.  The bottom-left plot is, again, similar to other plots
that we have seen.   The bottom-right plot is, however, of particular interest because the blue
dot-dashed curve in this plot vanishes:  in the local fluid rest frame to zeroth order in
gradients the drag force must be parallel to the direction of motion of the heavy
quark; without the effects of fluid gradients, there can be no perpendicular component.
We have also seen in Section \ref{sec:bj} that if the expansion were boost invariant 
then in the local fluid rest frame the drag
force on the heavy quark would still act parallel to the direction of motion of
the heavy quark even when the effects of fluid gradients are taken into account to
first order.  Therefore, the fact that the black dashed curve in the bottom-right
panel of Fig.~\ref{betax07betaz04} is nonzero is a direct manifestation of
the effects of fluid gradients {\it and} of the fact that the expanding fluid
produced in the collision of the two sheets of energy is not boost invariant.
The magnitude of the force described by this curve is small, since the
expansion is close to boost invariant, but it is nonzero.  We also see
that the first order effects of fluid gradients push the black dashed
curve toward the full result, shown as usual by the red curve.

We conclude from the investigations that we have reported in this section
that the discrepancies observed in Ref.~\cite{Chesler:2013cqa}
between the actual drag force on a heavy quark being pulled through
the matter produced in the collision of sheets of energy 
and the drag force that would have been obtained in an
static, homogeneous, plasma with the same energy
density is indeed due to the effects of spatial gradients
in, and time derivatives of, the fluid on the drag force.
Evaluating these effects to first order in the fluid gradients
explain all the qualitative aspects of the discrepancies
found in Ref.~\cite{Chesler:2013cqa} and do a reasonable
job even at the quantitative level.

\section{Future directions}
\label{sec:dis}

In (\ref{f_final}) we have derived a general expression for the
drag force needed to pull a heavy quark
through a dynamic fluid, 
flowing in some arbitrary fashion described by hydrodynamics,
to first order in the gradients and time derivatives of the fluid velocity.
We have applied this result to heavy quarks moving through
a fluid that is expanding according to Bjorken flow and to heavy
quarks moving through the expanding and cooling liquid produced
in a collision of sheets of energy in strongly coupled ${\cal N}=4$ SYM
theory.  Future directions include applying (\ref{f_final}) to heavy
quarks moving through strongly coupled plasma whose dynamics is
described by
other hydrodynamic solutions, for example including transverse
expansion.     It would also be interesting, and challenging, to extend (\ref{f_final})
to second order in fluid gradients.  Doing so could clarify how
the drag force behaves in the large $\gamma$ limit in the case
where, as we have done, one assumes that the quark mass $M\rightarrow\infty$
limit has been taken first.   
We have seen that in this regime when (\ref{limit_gamma}) is
not satisfied the first order contributions of the fluid
gradients dominate over the zeroth order drag force, which motivates
an evaluation of the magnitude of the second order contributions.
Considering the effects of fluid gradients on
a finite-mass quark at a large enough $\gamma$ that (\ref{old_limit_gamma}) 
is not satisfied would, however, require a different calculation
entirely.   The right starting point for this would be an analysis of
the rate of energy loss and transverse momentum broadening of 
a light quark in a dynamic strongly coupled fluid, including the effects
of fluid gradients.  
Other interesting directions would include investigating
how gradients in the fluid affect the emission of photons
and dileptons or the screening of the attraction between
a heavy quark and antiquark, and hence how they affect the binding
or dissociation of quarkonia. 
We leave the holographic calculation of the
effects of fluid gradients on these and other probes of the strongly
coupled fluid to future work.

\begin{acknowledgments}
We are particularly grateful to Paul Chesler,
with whom we collaborated
in doing the work reported in Ref.~\cite{Chesler:2013cqa}
that prompted the present study.  We had many very helpful
discussions with him as we began this work and throughout.
We would also like to thank 
Hong Liu and Navid Abbasi for helpful discussions.
This work was supported by the U.S. Department of Energy
under cooperative research agreement DE-FG0205ER41360.
\end{acknowledgments}

\end{document}